\documentclass[aps,prl,twocolumn,amsmath,amssymb,floatfix, showpacs]{revtex4-1}
\usepackage{graphicx}
\usepackage{dcolumn} 
\usepackage{bm}      
\usepackage[usenames,dvipsnames]{color}
\usepackage{ulem} 
 \usepackage{amsmath}
 \usepackage{amssymb}

\usepackage[T1]{fontenc}
\usepackage[utf8]{inputenc}
\usepackage{lmodern}
\usepackage{microtype}
\usepackage{amsmath}
\usepackage{epstopdf}
\usepackage{hyperref}
\usepackage{lipsum}
\usepackage{indentfirst}
\usepackage{latexsym}
\def \tauP {\tau^{\rm P}}

\def \f {{\bm f}}

\def \dive {{\bm \nabla}\cdot}

\newcommand{\deldel}[2]{\frac{\partial #1}{\partial #2}}

{}

\def \Ca  {\mbox{Ca}}
\def \Cb  {C_{\rm b}}

 \def \ac {a_{\rm c}}

\def \Gs {G_{\rm s}}
\def \Gb {G_{\rm b}}

\newcommand \bx {\bm{x}}

\def \rj {{\rm j}} 
\def \ff {\bm{f}}
\def \xj {\bm{x}_{\rm j}}
\def \lone {\lambda_{1}}
\def \ltwo {\lambda_{2}}
\def \Ione {\mathcal{I}_{1}}
\def \Itwo {\mathcal{I}_{2}}

\newcommand\bu{\ensuremath{\mathbf{u}}}

\usepackage{graphicx,bm,ulem,dcolumn}
\usepackage[usenames,dvipsnames]{color}

\begin{document}
\title{Motion of an elastic capsule in a constricted microchannel}
 \author{Cecilia Rorai$^{1,2}$, Antoine Touchard$^{1,3}$, Lailai Zhu$^{1,4}$ and Luca Brandt$^{1}$}
 \affiliation{$^1$Linn\'e Flow Centre and Swedish e-Science Research Centre (SeRC), KTH Mechanics, SE-10044 Stockholm, Sweden;\\
 $^2$ENSTA ParisTech, Palaiseau, France;\\
 $^3$Nordita, Roslagstullsbacken 23, 106 91 Stockholm, Sweden.}


\begin{abstract}
We study the motion of an elastic capsule through a microchannel characterized by a localized constriction. We consider a capsule with a stress-free spherical shape and impose its steady state configuration in an infinitely long straight channel as the initial condition for our calculations.
We report how the capsule deformation, velocity, retention time, and maximum stress of the membrane are affected by the capillary number, $Ca$, and the constriction shape.
We estimate the deformation by measuring the variation of the three-dimensional surface area and a series of alternative quantities easier to extract from experiments. These are the Taylor parameter, the perimeter and the area of the capsule in the
spanwise plane. We find that the perimeter is the quantity that reproduces the behavior of the three-dimensional surface area the best. This is maximum at the centre of the constriction and shows a second peak after it, whose location depends on the $Ca$ number. We observe that, in general, area-deformation correlated quantities grow linearly with $Ca$, while velocity-correlated quantities saturate for large $Ca$ but display a steeper increase for small $Ca$.
The velocity of the capsule divided by the velocity of the flow displays,
surprisingly, two different qualitative behaviors for small and large capillary numbers.
Finally, we report that longer constrictions and spanwise wall bounded (versus spanwise periodic) domains cause larger deformations and velocities.
If the deformation and velocity in the spanwise wall bounded domains are rescaled by the initial equilibrium deformation and velocity, their behavior is undistinguishable from that in a periodic domain. In contrast, a remarkably different behavior is reported in sinusoidally shaped and smoothed rectangular constrictions indicating that the capsule dynamics is particularly sensitive to abrupt changes in the cross section. In a smoothed rectangular constriction larger deformations and velocities occur over a larger distance.
\end{abstract}
\maketitle

\section{Introduction }

Capsules, closed fluid-filled membranes of elastic material, are a widely accepted model to simulate the hydrodynamical behavior of living cells \cite{freund_13_review} as well as manufactured microcapsules and microspheres.
Theoretical and numerical investigations of the dynamics of deformable capsules in confined flows are motivated by a better understanding of physiological systems ({\it e.g.}\ the flow in vascular capillaries), and by designing devices for medical diagnostic \cite{mao,limhoo14}, pharmaceutical, food and cosmetic industry. In this study we analyze specifically the case of a deformable capsule transported through a channel characterized by a localized constriction. 

Constrictions are common in physiological systems and their length scales
span a wide range. Stenoses form in the arterial vessels due to the accumulating plague 
significantly narrowing the vessels~\cite{berger2000flows}. 
Red blood cells (RBCs) go through the slits in the spleen that are smaller than $1\mu$m. 
Such tiny constrictions can effectively trap stiff aged 
RBCs~\cite{mokken1992clinical,del2012role}, and malaria-affected RBCs, which are less deformable than the
 healthy ones due to the presence of the parasite; the trapped cells will be attacked by the
immune system. In the meantime, the cells going through the splentic slits are severely deformed
and subject to significant hydrodynamic and mechanical stresses, which also
serve to get rid of parasites without destroying the infected cells (a process called
'pitting'~\cite{schnitzer1972pitting,del2012role}).  

Constrictions are also utilized in micro-fluidic devices to manipulate cells and their 
synthetic counterparts. Experiments~\cite{shelby2003microfluidic} using confined elastomeric
constrictions inside microchannels have been performed to characterize the complex behaviour of healthy and
\textit{Plasmodium falciparum}-infected erythrocytes. Smartly designed constrictions are introduced
into microfluidic devices to differentiate and sort cells by deformability~\cite{bow2011microfabricated,qi2012probing,zhu2014_sorting} by exploiting the concept
that  a more floppy cell passes through the constriction faster than a stiff one, or that different deformations lead to different trajectories \cite{zhu2014_sorting}.
In a recent study~\cite{wyss2010capillary}, soft particles are driven into a 
constricted microcapillary by hydrodynamical pressure, and their elastic modulus is measured from their deformation.

We remark that deformation through a constriction is also a key element to understand how Deterministic Lateral Displacement (DLD) type of devices \cite{huang04} work when used to sort by deformability \cite{lund}. In fact, the flow between two obstacles, the basic building block of DLD devices, is in all respects similar to the flow through a constriction. This problem is also interesting for drug delivery applications, because it is through deformation in a shear flow that drugs are trapped in a carrier and then released. Erythrocytes themselves are good candidates as carriers \cite{Muzykantov10,Favretto13}. The majority of the methods for the entrapment of chemicals, drug, proteins, etc. in erythrocytes takes advantage of their remarkable capacity for reversible shape changes and for reversible deformation under stress, allowing transient opening of pores large enough to be crossed by externally placed macromolecules.

Given this range of applications, our objective is twofold: ({\it i}) characterize a single capsule behavior in a constriction and ({\it ii}) determine the relevant quantity to measure to most efficiently sort by deformability. In other words, we ask whether it is the velocity, deformation or stress the most characterizing observable for the behavior of a capsule flowing through a constriction. This second goal is relevant if our simple device is used to measure the capsule properties or sort by deformability. Indeed, sorting can be achieved by taking images of the cell in a constriction, processing them, and opening downstream gates according to the deformability \cite{Pra03}.

Park et al.  \cite{Park13} investigated the transient dynamics of an elastic capsule flowing in a square microchannel with a rectangular constriction and have compared it to that of a droplet: these authors note that the confinement and expansion dynamics of the fluid flow results in a rich deformation behavior for the capsule, from an elongated shape at the constriction entrance, to a flattened parachute shape at its exit. In a more recent study, Kuster et al. address the question of how the constriction width, container mechanics and external forcing affect the capsule dynamics and determine whether it will get trapped into the constriction or pass through it and how fast \cite{kusters14}.

As a last remark, note that there is a clear mathematical distinction between capsules and vesicles. Vesicles conserve surface area and volume and resist bending, whereas capsules conserve volume only and resist elastic shear (resistance to bending can be added to the model). Since capsules resist shear elasticity they are regarded as an adequate model for artificial polymerised capsules and RBCs, whose membrane is made up of a lipid bilayer coated form inside by a network of spectrin filaments. In \cite{noguchi05}, it is shown that the resistance to shear is responsible for qualitatively different behaviours in capillary flow (for example preventing the prolate to oblate shape transition). 
We believe that when studying the motion through narrow constrictions resistance to shear should be taken into account given the large viscous shear stresses that induce large deformations. Nevertheless there is a wealth of numerical studies on the motion of vesicles in shear flows which led to the observation of discocytes and slippers shapes in a Poiseuille flow~\cite{Danker09,kaoui2009red,kaoui2011complexity}.

Our paper is organized as follows. First, we describe the mathematical models and the numerical method used to
tackle 
the problem. Second, we describe the flow geometry and the initial conditions
employed in our calculations. Finally we present our results, focusing in particular on the effect of
the capillary number and of different geometries,~{\it~i.~e.},~longer and asymmetrical
constrictions. The paper ends with a summary of the main findings.



\section{Models and geometry}
\subsection{Mathematical model}

Our study is motivated in the context of microfluidic devices and is relevant to applications in medical diagnostic  that involve mechanical characterization and sorting  of cell samples. Typical cell sizes measure few tens of micrometers, while microfluidic devices designed to manipulate them have widths of the same order of magnitude, and characteristic velocities between few $\mu m/s$ to few $cm /s$ for single file cell devices \cite{lund,Kim09,Xuan10}.  Given the small sizes and velocities involved, it is justified to model the fluid flow by neglecting the inertial effects and by reducing the Navier-Stokes equations to the linear Stokes equations. 

In our study, capsules are regarded as fluid-filled closed membranes of elastic material that are assumed to be two-dimensional and isotropic. Their deformations are measured as the displacements from a reference shape that is assumed to be spherical with radius $\ac$ (note that the debate is open on the possibility that a  stress-free spherical shape is a correct descriptions for red blood cells \cite{peng14,cordasco14}). We have chosen to describe the capsule through a neo-Hookean constitutive law according to which the local strain energy function is

\begin{equation}
W=\frac{\Gs}{2}\left[\Ione -1 + \frac{1}{\Itwo+1}\right],
\label{eq:nHookean}
\end{equation}
where $\Gs$ is the isotropic shear modulus, while $\Ione = \lone^2+\ltwo^2-2$ and $\Itwo=\lone^2\ltwo^2-1$ 
are the two invariants of the left Cauchy-Green tensor expressed in terms of the eigenvalues, $\lone$, $\ltwo$, of the two-dimensional strain matrix, also referred to as the principal stretch ratios. This model has been often used to describe biological membranes, it is appealing because of its relative simplicity which allows us to better identify the interplay between the viscous and elastic forces.

We also employ a linear isotropic model for the bending moment~\cite{zhao10,lai14},
with a bending modulus $\Gb= \Cb \ac^2 \Gs$, where  
$\Cb = 0.01$ is held constant in our simulations; this value is consistent with available experimental data for
RBCs~\cite{freund_13_review}. In our calculations we have verified that for a capsule that undergoes large deformations in the constriction ($Ca=0.6$) the shearing energy is about ten times larger than the bending energy.
Finally we also assume that the fluid inside and outside the cell
has exactly the same density and viscosity.

The surface of the capsule is discretized by $N$ points and moves with the flow obeying no-slip, non-penetrating boundary conditions. The velocity field displaces the surface of the capsule out of its equilibrium configuration causing a back-reaction of the membrane to the flow that enters the flow equation as a forcing term $\sum_{\rj=1}^{N} \f_{\rj}\delta \left(\bx -\xj \right)$ at the collocation points $\xj$. Hence the overall flow field results from the superposition of responses induced by each single point force (Green functions) acting on the surface of the capsule and on the walls. 

The full set of equations includes the three-dimensional Stokes equation, the continuity equation for an incompressible flow and the equation to evolve the surface of the capsule  

\begin{eqnarray}
\label{eq:stokes}
-\nabla p + \mu \nabla^2 \bu & = & -\sum_{\rj=1}^{N} \f_{\rj}\delta \left(\bx -\xj \right),  \\ 
\label{eq:divu}
\dive \bu &  =  &  0,
\end{eqnarray}

\begin{equation}
\frac{d \xj}{dt} = \bu(\xj)\/. 
\end{equation}
Here $p$ is the pressure, $\bu$ is the velocity, and $\mu$ is the dynamic viscosity of the fluid. 

The elastic stresses on the surface of the capsule are given by the
two principal tensions, $\tauP_1$ and $\tauP_2$, defined by:~\cite{skalak1973strain}
\begin{eqnarray}
\tauP_{1} &=& 2\frac{\lone }{\ltwo} 
        \left[ \deldel{W}{\Ione}+\ltwo^{2}\deldel{W}{\Itwo} \right], 
               \nonumber \\
  \tauP_{2} &=& 2\frac{\ltwo}{\lone}
     \left[ \deldel{W}{\Ione}+\lone^2\deldel{W}{\Itwo} \right].
\label{eq:tension}
\end{eqnarray}

The dimensionless number that characterizes this problem is the capillary number,
$\Ca  \equiv \mu U/\Gs$, where U indicates the average flow velocity. The capillary number expresses the ratio between the viscous forces and the elastic forces. Given the same flow conditions, higher capillary numbers refer to softer capsules.  

\subsection{Numerical methods}

Equations (\ref{eq:stokes}) and (\ref{eq:divu}) are solved by a hybrid boundary Integral-Mesh
method, the General Geometry Ewald like method (GGEM)~\cite{kumar12,graham07_prl,freund_13_review}, used in a variety of micro-multiphase simulations \cite{kumar11,pranay10,pranay12}. 
In our implementation, the mesh-based part (responsible for the
long-range part of the Green's function) is calculated by the spectral-element solver
NEK5000~\cite{nek5000-web-page} which allows us to cope with non-trivial boundaries.
The short-range part is handled by standard boundary integral techniques. 

The advantage of the GGEM method   
is the following. If $N$ is the number of singular point forces present in the flow, building the mobility matrix according to a traditional boundary integral method (BIM) requires $N^{2}$ operations. The total computational cost will have also to include the operations needed to 
solve the corresponding linear system. The quadratic scaling comes from the fact
that the traditional BIM considers the hydrodynamical interactions between every pair 
of point forces, independently of their distance. This scaling makes the BIM prohibitive to study problems that involve a large number of particles. In the GGEM scheme, instead, the number of operations for the local solution scales 
linearly with $N$, whereas the scaling of the global problem depends on the mesh-based solver, as well as the geometry and boundary conditions of the computational domain. The long-ranged interaction
is here solved by a highly parallel Stokes solver on an Eulerian grid.

We use a spectral discretization based on spherical harmonics as in~\cite{zhao10} to calculate $\ff_{\rj}$ given the
positions $\xj$, i.e.\ to solve the membrane stress balance. The shape of the capsule is mapped onto a unit sphere. In most cases, we use $25$ and 
$50$ modes along the latitudinal and longitudinal direction, resulting in a total of $1250$ discretized points. We have 
shown in ~\cite{lai14} that by using $24 \times 48$ modes, we get excellent agreement for a capsule tightly squeezed in 
a square-duct, where the diameter $D$ of the undeformed capsule over the size $l$ of duct is $D/l=0.9$.
A more detailed description of the numerical algorithm can be found in ~\cite{lai14,lailai_phd}.  

\subsection{Flow geometry and boundary conditions}

\begin{figure}[!htbp]
\centering
{\includegraphics[width=.8\linewidth]{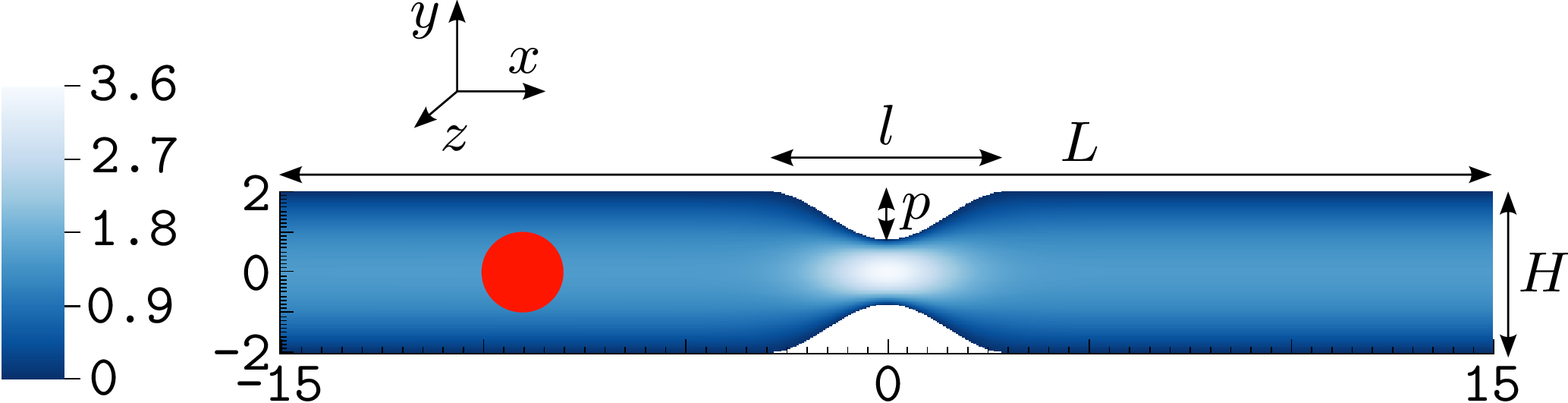}}
\put(-215,37){{\large (a)}}\\
{\includegraphics[width=.8\linewidth]{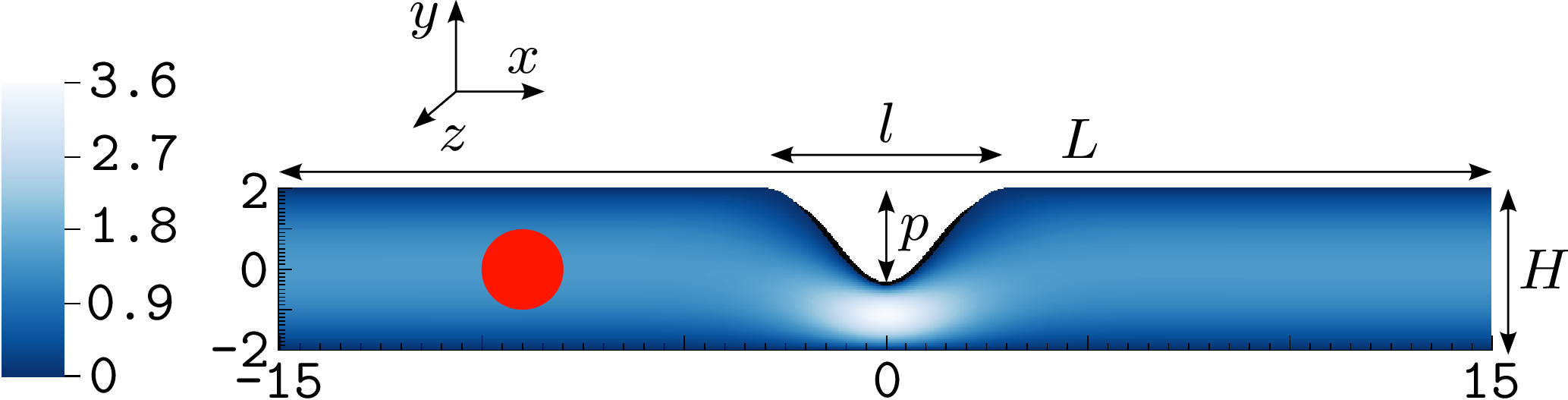}}
\put(-215,37){{\large (b)}}\\
{\includegraphics[width=.8\linewidth]{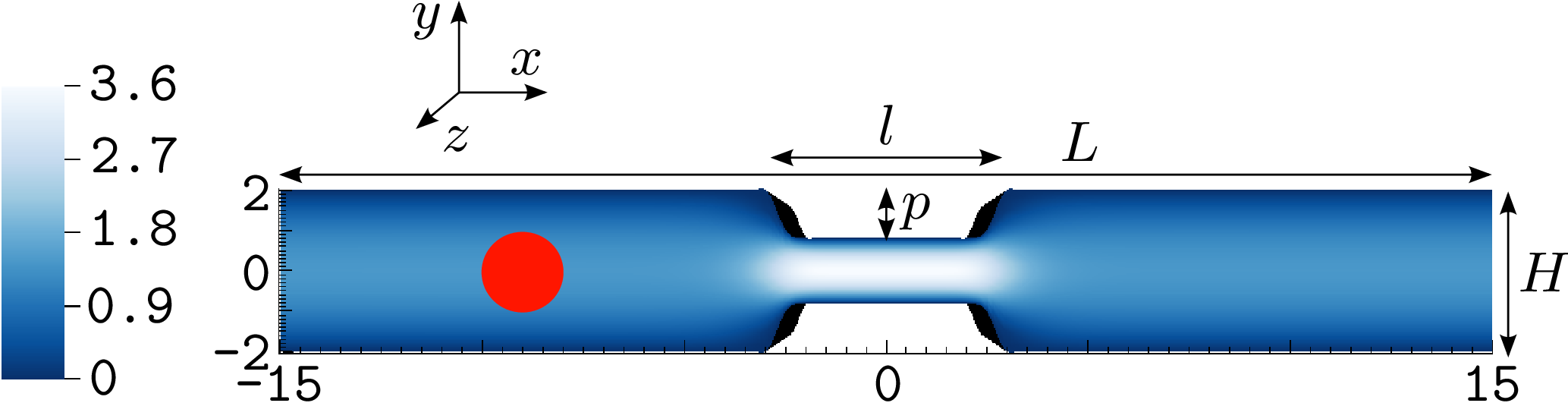}}
\put(-215,37){{\large (c)}}\\
\caption{Sketch of the various computational domains: channels with different geometrical
constrictions, shown on the $z=0$ plane. (a) Sinusoidal  symmetric  constriction, (b) sinusoidal asymmetric constriction; and (c) rounded rectangular constriction. The colormap indicates the magnitude of the flow velocity in the absence of the capsule. The undeformed shape of the capsule of radius $a_c=1$, and its initial position
are indicated by the red circle.} 
\label{fig:sketch}
\end{figure}

We investigate the motion of a three-dimensional elastic capsule passing through a constricted microchannel (Figure~\ref{fig:sketch}). The stress-free shape of the capsule is a sphere of radius $\ac$. The channel is oriented in the $x$ direction and is vertically bounded by walls in the $y$ direction. Its height is
$H=4\ac$. Along the spanwise $z$ direction, whose depth is $D$, either periodic ($D=14\ac$), or no-slip ($D=4\ac$) 
boundary conditions are imposed. The latter corresponds to a flow in a square duct.

A velocity flow profile is imposed at the inlet, no slip and no penetrating boundary conditions are imposed on the solid walls and zero-stress boundary conditions, $-p\mathbf{I}+\mu\left(\nabla\mathbf{u}+\left(\nabla \mathbf{u}\right)^{T}\right)=0$, are 
imposed at the outlet. The inlet velocity flow profile is the parabolic  plane Poiseuille profile for periodic boundary conditions in $z$, and the solution for a square duct from \cite{spiga1994symmetric} otherwise.

We have studied the motion through a sinusoidal [Figure~\ref{fig:sketch}(a)-(b)], and a rounded rectangular [Figure~\ref{fig:sketch}(c)] constriction; for the former case an  asymmetric [Figure~\ref{fig:sketch}(b)] setup with respect to the $x$-axis is  also
considered.
We have run simulations for different
values of the constriction length, $l$, while the constriction width, $h$, is kept constant: $h=1.6$.

The channel walls are located at $y_{w}=\pm 2$, and the
constriction, centered at $x=0$, extends between $-l/2<x<l/2$. The constriction wall $y_{con}$ is give by a trigonometric function:
\begin{eqnarray}
y_{con}=(y_{w}+\bar{y})\frac{h}{H}+(&y_{w}&+\bar{y})\left(1-\frac{h}{H}\right)\sin^d\left(\pi\frac{x}{l}\right)-\bar{y},\nonumber \\
 &x&\in\left[-\frac{l}{2},\frac{l}{2}\right], \label{equ:mesh_deformation}
\end{eqnarray}
where $l$, $\bar{y}$ and $d$ are parameters. We have performed calculations for a constriction length $l=4,6,8$. For the symmetric and asymmetric case $\bar{y}=0,2$ respectively, while $d$ controls the constriction shape and it is $d=2$ for the sinusoidal constriction, and $d=48$ for the ``rounded rectangular'' constriction. 

The runs in the symmetric sinusoidal constriction are named by ``S'', those in the asymmetric sinusoidal constriction by ``A'', calculations in the rounded rectangular constriction are denoted by ``R'' and those in a duct with a symmetric sinusoidal constriction by ``D''. In the following, numbers following capital letters, as found in Table \ref{tab:channel_dimension_sym}, indicate the length of the constriction. The dimensions of the channel, including the parameters used for the constriction shape for all the simulations reported in this paper, are listed in Table \ref{tab:channel_dimension_sym}. 
The velocity magnitude of the flow in the
absence of the capsule is also displayed in Figure~\ref{fig:sketch}.

\begin{table*}
\begin{center}
\begin{tabular}{|p{4.5cm} l |*{6}{c|}}
\hline
\centering
\bf Geometry                &          & \bf S4      & \bf S6             & \bf S8         & \bf A6      & \bf R6      & \bf D6         \\
\hline
\bf Channel length          & $L$      & 28          & 30                 & 32             & 30          & 30          & 30             \\
\hline
\bf Channel depth           & $D$      & 14          & 14                 & 14             & 14          & 14          & 4              \\
\hline
\bf Constriction length     & $l$      & 4           & 6                  & 8              & 6           & 6           & 6              \\
\hline
\bf Constriction position   & $p$      & 1.2         & 1.2                & 1.2            & 2.4         & 1.2         & 1.2            \\
\hline
\bf Symmetry parameter     & $\bar{y}$ & 0           & 0                  & 0              & 2           & 0           & 0              \\
\hline
\bf Constriction shape      & $d$      & 2           & 2                  & 2              & 2           & 48          & 2              \\
\hline
\end{tabular}
\end{center}
\caption{Dimensions of the microchannel for the sinusoidal symmetric (S), 
the sinusoidal asymmetric (A) and the smoothed rectangular (R) constriction geometries, plus a symmetrically and sinusoidally constricted duct domain (D). The domain geometry, dimensions, and notation are sketched in Figure \ref{fig:sketch} and discussed in the text. Several runs are
performed in each of these geometries for a variety of capillary numbers ranging between $Ca=0.05$ and $Ca=0.6$.}
\label{tab:channel_dimension_sym}
\end{table*}

\section{Results} \label{sec:results}
We firstly calculate the steady shape of a capsule in an infinitely long channel with spanwise periodic boundary conditions or side walls. Simulations are performed for capillary numbers ranging between $Ca=0.05$ and $Ca=0.6$. Detailed results are reported only for $Ca=0.05, 0.3, 0.6$, for clarity. 
The equilibrium shape is used as initial condition for calculations in the constricted channels.

In most of our computations, the flow is assumed to be periodic in the spanwise direction and we investigate the effect of the capillary number and the constriction geometry (length, shape, and asymmetry, see Figure \ref{fig:sketch}) on the transient dynamics of the capsule. We later consider the motion in a spanwise-confined duct for comparison. We characterize the dynamics in terms of the deformation of the capsule, velocity of the centroid, and maximum stresses on the membrane.  Various estimates of the capsule deformation are proposed to facilitate a direct comparison with experimental results. 
\subsection{Capsule steady state in a straight channel}
\begin{figure}[!htbp]
\begin{center}
\resizebox{5.5cm}{!}{\includegraphics[width=\linewidth]{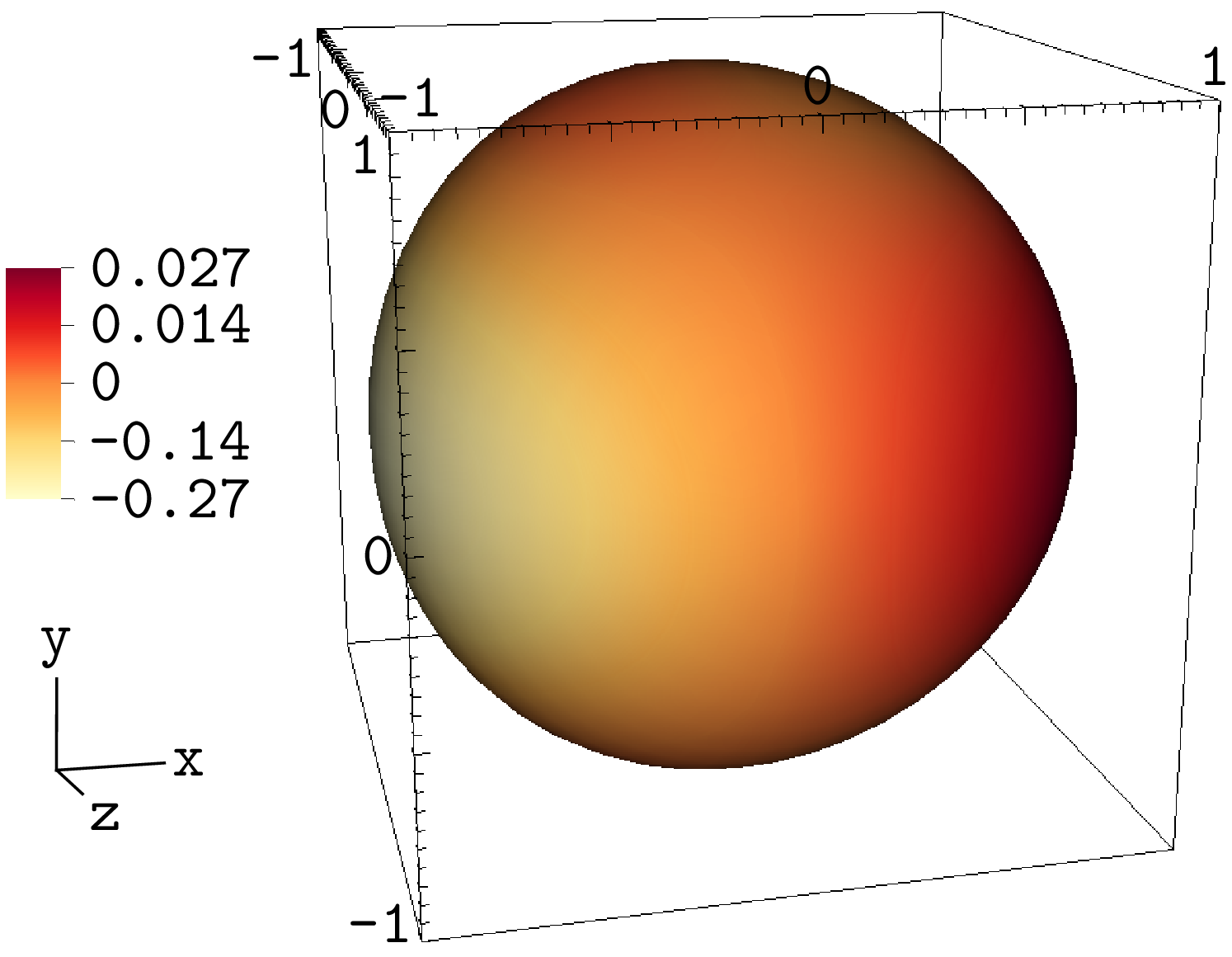}}
\put(-200,110){{\large (a)}}\\
\resizebox{5.5cm}{!}{\includegraphics[width=\linewidth]{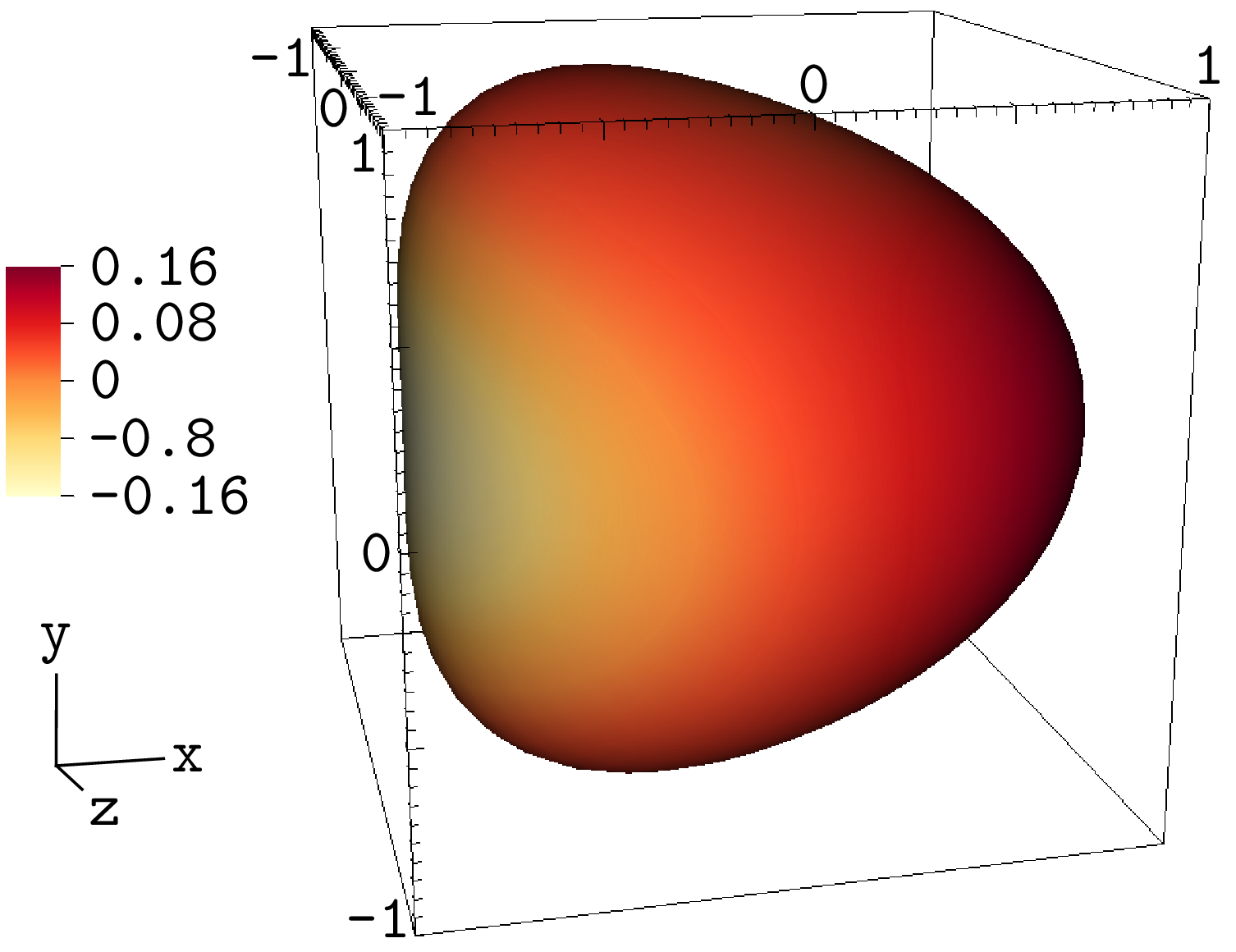}}
\put(-200,110){{\large (b)}}\\
\resizebox{5.5cm}{!}{\includegraphics[width=\linewidth]{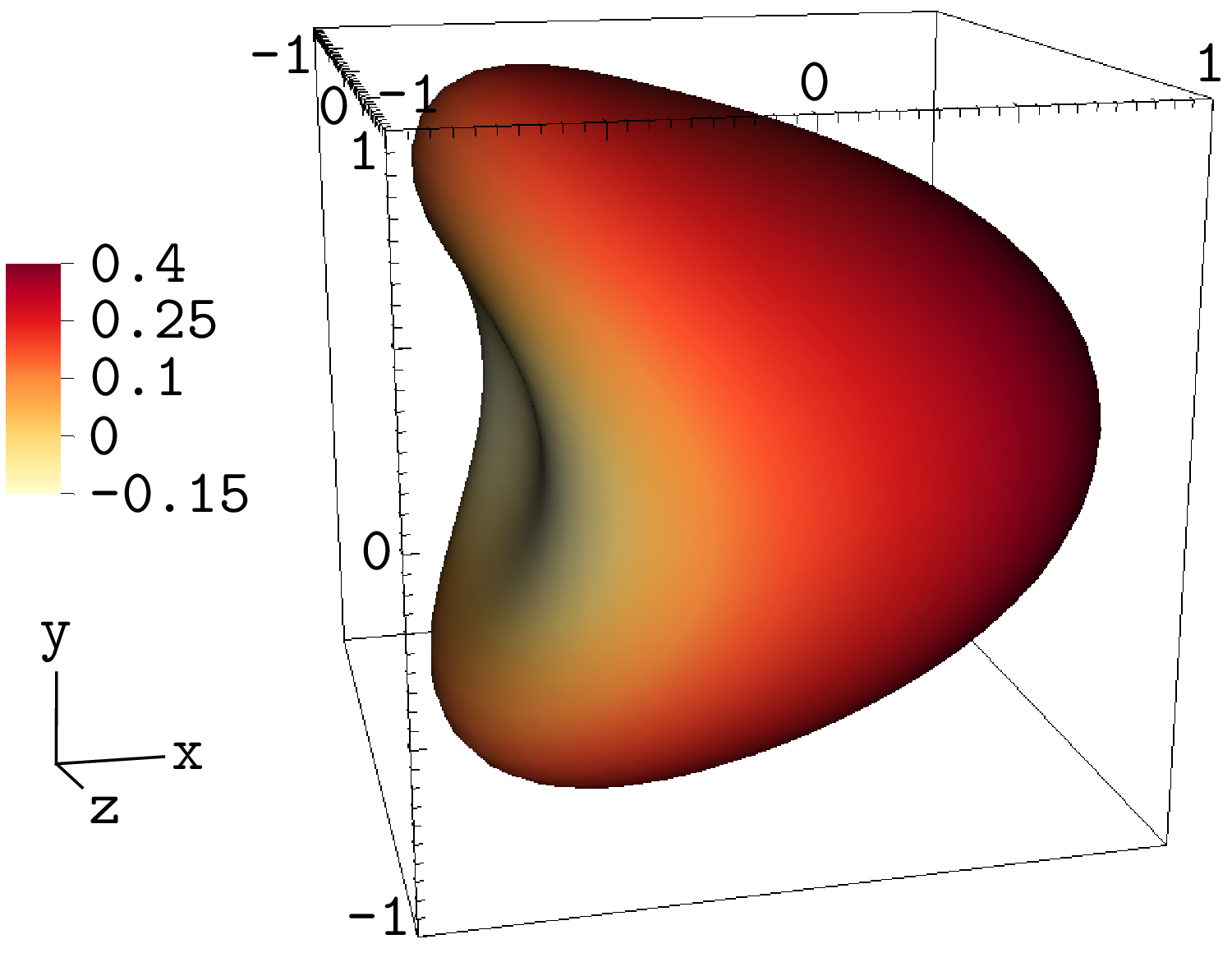}}
\put(-200,110){{\large (c)}}
\end{center}
\caption{Steady state shapes and magnitude of the principal isotropic stress (in color) for the motion in a straight channel of capsules with different capillary numbers: (a) $Ca=0.05$, (b) $Ca=0.3$ and (c) $Ca=0.6$. The channel is 4$\ac$ unit high and is periodic in the spanwise direction. As the capillary number increases the capsule develops a bullet like and a croissant like shape. The stress is larger on the front part, and this is where rupture may occur.}
\label{fig:steady_state}
\end{figure}
\begin{figure}[!htbp]
\centering
\resizebox{6.5cm}{!}{\includegraphics[width=\linewidth]{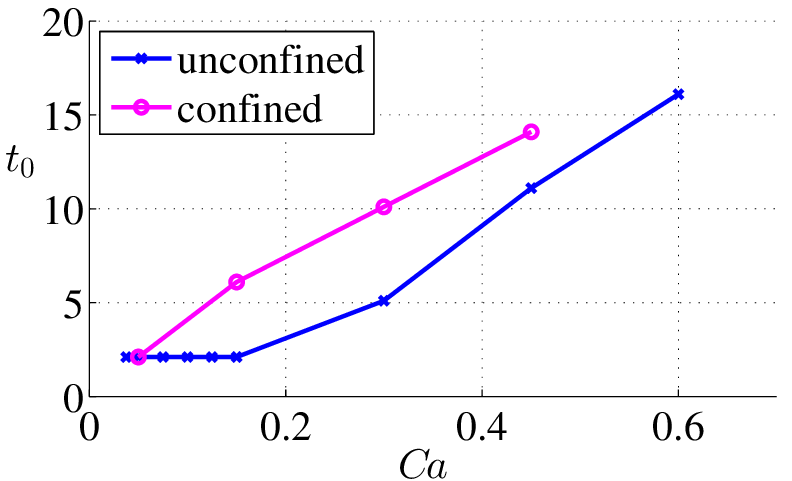}}
\caption{Dimensionless time needed to converge to the steady state as a function of the capillary number. Two different spanwise boundary conditions are considered as listed in the legend: unconfined or $z$-periodic boundary condition, and $z$-confined boundary condition correspondent to the flow in a square duct.}
\label{fig:steady_state_des}
\end{figure}
\begin{figure}[!htbp]
\centering
\resizebox{6.5cm}{!}{\includegraphics[width=\linewidth]{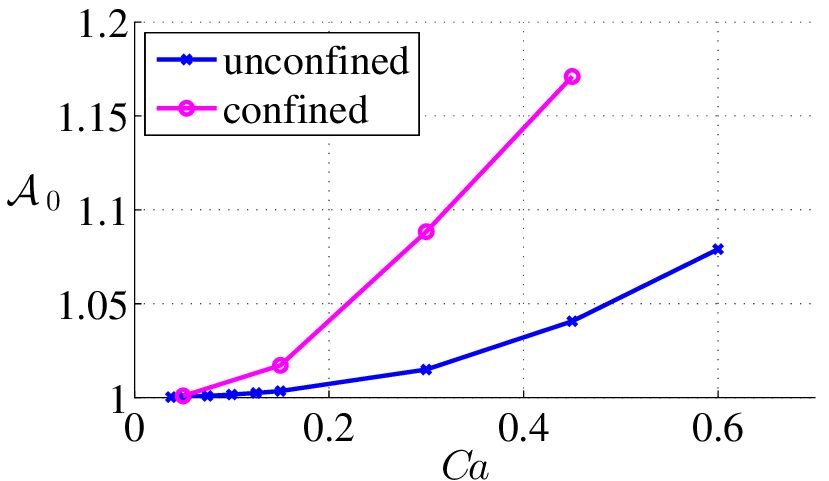}}
\put(-200,95){{\large (a)}}\\
\resizebox{6.5cm}{!}{\includegraphics[width=\linewidth]{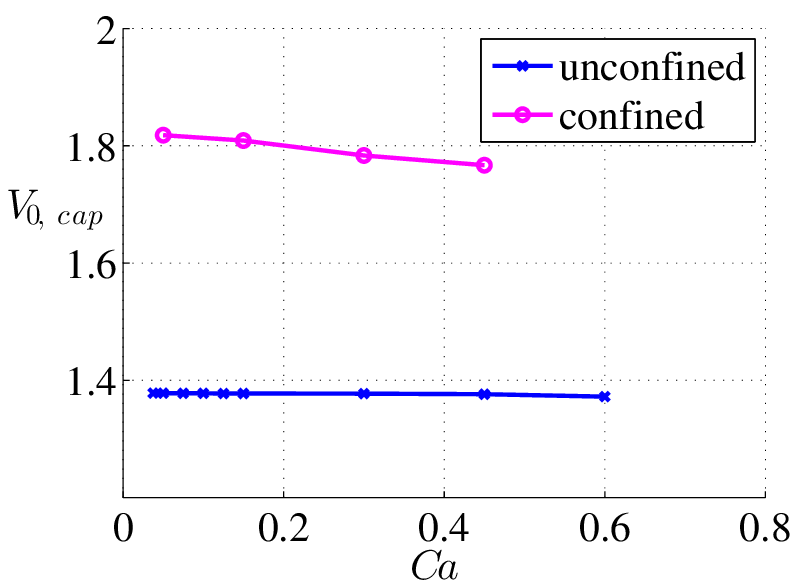}}
\put(-200,115){{\large (b)}}\\
\resizebox{6.5cm}{!}{\includegraphics[width=\linewidth]{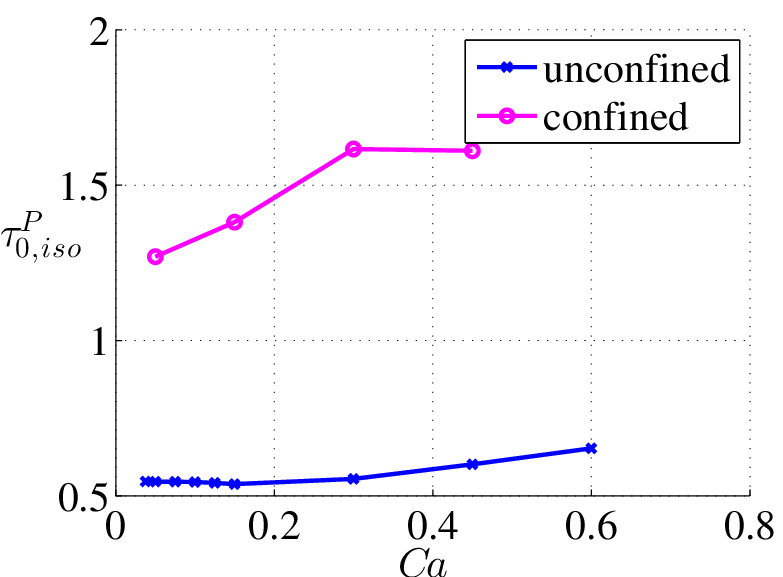}}
\put(-200,115){{\large (c)}}
\caption{Value of (a) the steady state fractional change in the area, $\mathcal{A}_0$, (b) velocity of the capsule, $V_{0, cap}$, and (c) principal isotropic stress, $\tau^P_{0, iso}$,  as a function of the capillary number. We compare the unconfined domain (periodic boundary conditions in $z$) with the confined case (square duct) as reported in the legend. These results do not depend on the constriction geometry since they are calculated for a straight channel/duct.}
\label{fig:steady_state_des2}
\end{figure}
The center of mass of the capsule is initially located on the channel mid-line ($y=0$, $z=0$), as shown in Figure \ref{fig:sketch}. 
Capsules are believed to migrate towards the centerline if initially offset, which has been systematically observed by us and other groups \cite{Doddi08,Rorai15}. However,  a careful analysis showed that this is not the case for vesicles \cite{Farutin14}, questioning previous believes and suggesting that a more systematic investigation should also be undertaken for capsules.
In a simple straight channel the mutual interaction between the flow and the membrane displaces the latter from its relaxed shape and deforms it into a steady state configuration that depends on the capillary number and the channel width \cite{kuriakose13}. We show the steady state deformation in Figure \ref{fig:steady_state} for $Ca=0.05, 0.3, 0.6$. The steady state is nearly spherical for stiff capsules, whereas, by increasing the capillary number, it develops a front-rear asymmetry and displays first a bullet-like, and then a croissant-like shape \cite{Kuriakose11,Coupier12,Farutin14}. We refer to the shape as ``croissant'' rather than ``parachute'' following the convention in \cite{Coupier12,Farutin14} according to which a ``parachute'' shape is perfectly axisymmetric. This is not the case here since the capsule either sees the periodic boundary conditions in $z$ and the walls in $y$ or adapts to the duct square cross-section \cite{Kuriakose11,Hu12}.

For convenience, we directly perform the steady state calculation in a long enough constricted domain: the initially spherical capsule is centered at the point $(-8,0,0)$, and is held fixed at that location until the steady state is achieved.
The convergence is set by a tolerance parameter, $toll=1e^{-3}$, imposed on the area variation: $$|A(t+1)-A(t)|<toll,$$ where $t$ is the dimensionless time expressed in units of $t_c =a_c/U$, the characteristic time scale. 

The capsule is anchored $5\ac$ away from both the inlet and the constriction, as shown in Figure \ref{fig:sketch}. 
It has been verified that by doing so, the inlet is not affected by the capsule, and the capsule in not affected by the flow pattern in the constriction. The capsule is released once the steady state is reached. 

The choice of using the steady state shape as initial condition guarantees that the subsequent transient dynamics is solely determined by the presence of the constriction, as opposed to being a combination of the convergence to a steady state and the dynamics in the constriction. 

In Figure \ref{fig:steady_state_des} the time needed to converge to the steady state is plotted as a function of the capillary number. Softer and more confined capsules take longer time to reach the equilibrium.
The steady state area deformation, centroid velocity, and local maximum isotropic stress  for different capillary numbers are plotted in Figure \ref{fig:steady_state_des2}(a)-(c). 

The parameter expressing the area deformation $\mathcal{A}_{0}$ is defined as the ratio between the steady state surface area $A_0$ and the stress-free or relaxed initial area, $A_{R}$, which is the area of a unit radius sphere,
\begin{align}
\mathcal{A}_{0} = \frac{A_0}{A_{R}}. \label{equ:surfaceareaexpansion}
\end{align}
The ``0'' index denotes the steady state condition.
Capsules with larger capillary numbers deform more and tend to elongate along the central region of the flow
 where the velocity is higher. The area $\mathcal{A}_{0}$ increases up to a $5\%$ in our range of $Ca$ for the periodic domain, and up to a $15\%$ in the duct. The minimum distance from the wall is almost unchanged due to the bulging of the rear part for high $Ca$; as a consequence, the velocity of the capsule does not display an appreciable variation, especially in the unconfined case [Figure \ref{fig:steady_state_des2}(b)]. The velocity slightly decreases with $Ca$.

The steady state principal isotropic stress, $\tau^{P}_{0, iso}$, is defined as the local maximum of the isotropic principal tension. Hence, if $\tau^P_1$ and $\tau^P_2$ are the two components of the stress,
\begin{align}
\tau^{P}_{0, iso} = \max_{\boldsymbol{x}} \left[\frac{\tau^P_1(\boldsymbol{x})+\tau^P_2(\boldsymbol{x})}{2}\right].
\end{align}
This parameter grows with the capillary number, especially in the confined case as seen in Figure \ref{fig:steady_state_des2}(c).
\begin{figure*}[htbp]
\resizebox{4cm}{!} {\includegraphics[width=\linewidth]{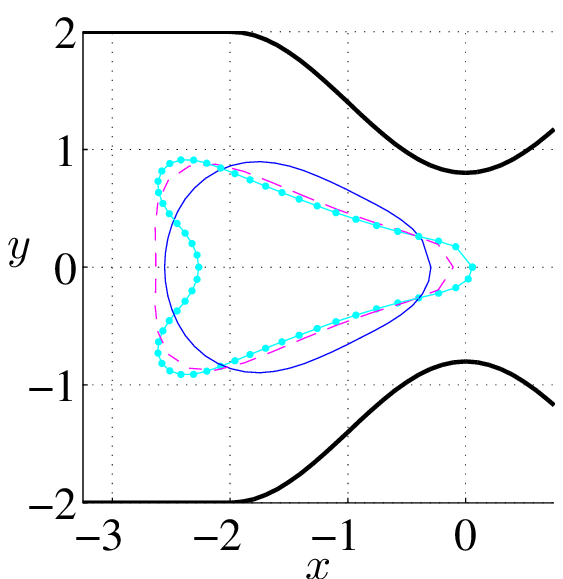}}
\put(-90,95){{\large (a)}}
\resizebox{4cm}{!}  {\includegraphics[width=\linewidth]{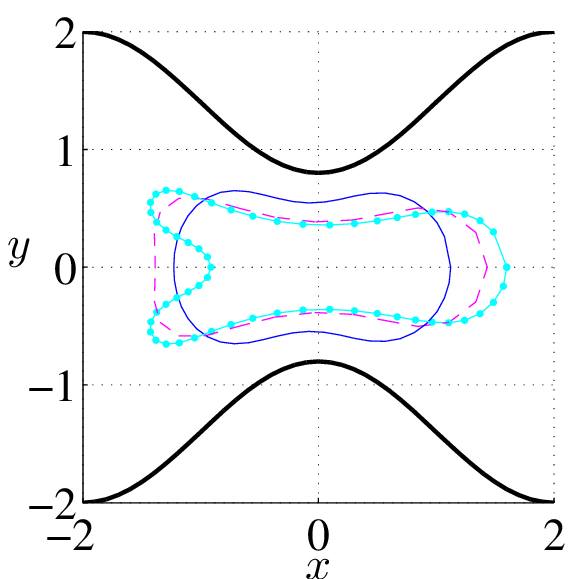}}
\put(-70,95){{\large (b)}}
\resizebox{4cm}{!}  {\includegraphics[width=\linewidth]{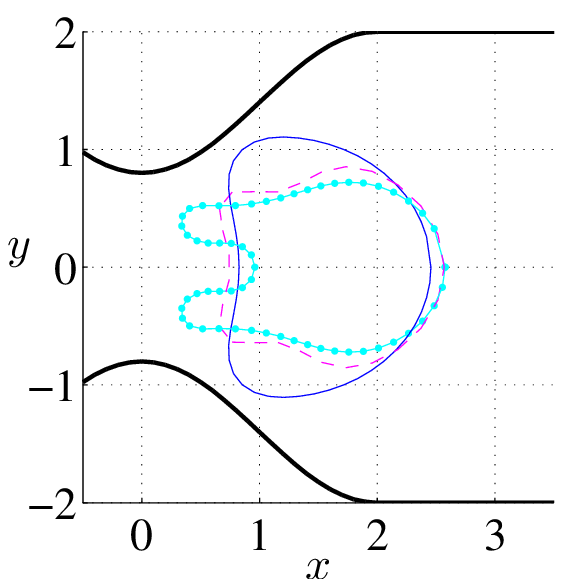}}
\put(-90,95){{\large (c)}}
\resizebox{4cm}{!}  {\includegraphics[width=\linewidth]{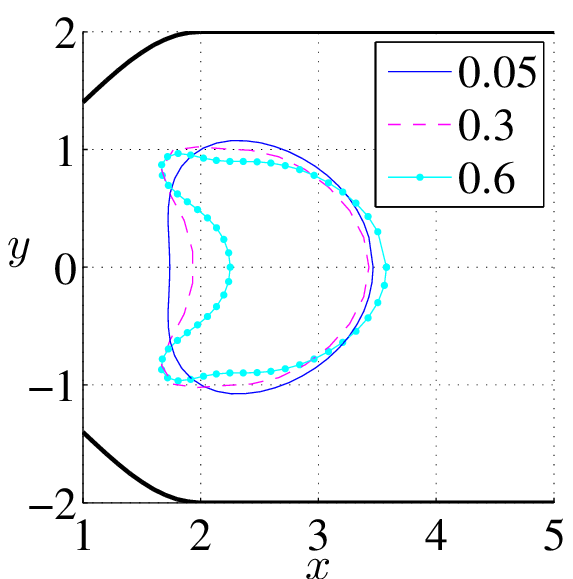}}
\put(-70,95){{\large (d)}}\\
\resizebox{4cm}{!}  {\includegraphics[width=\linewidth]{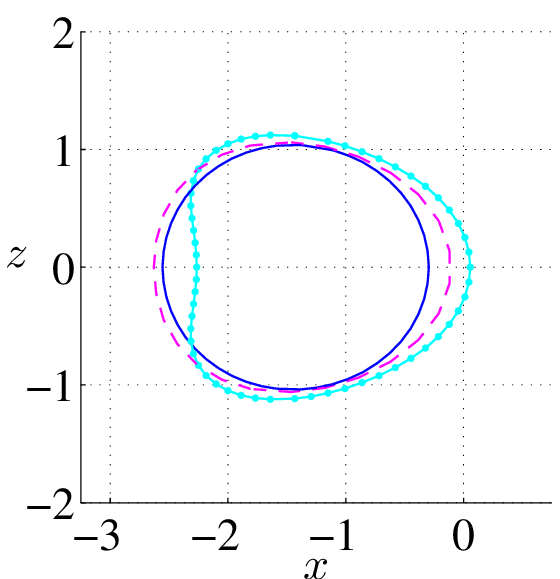}}
\put(-90,95){{\large (e)}}
\resizebox{4cm}{!}  {\includegraphics[width=\linewidth]{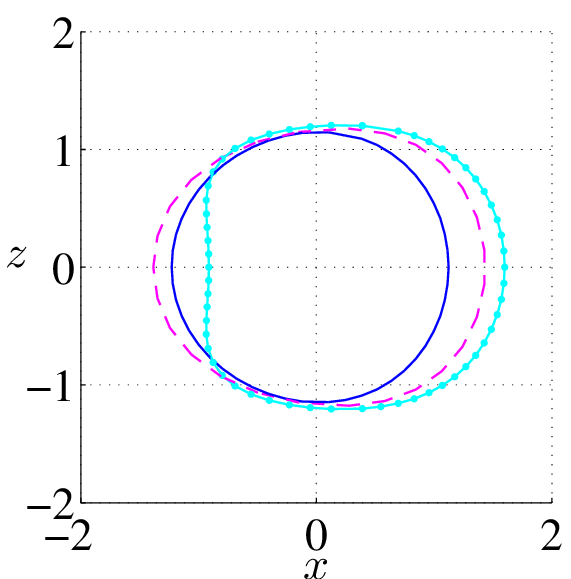}} 
\put(-90,95){{\large (f)}}
\resizebox{4cm}{!}  {\includegraphics[width=\linewidth]{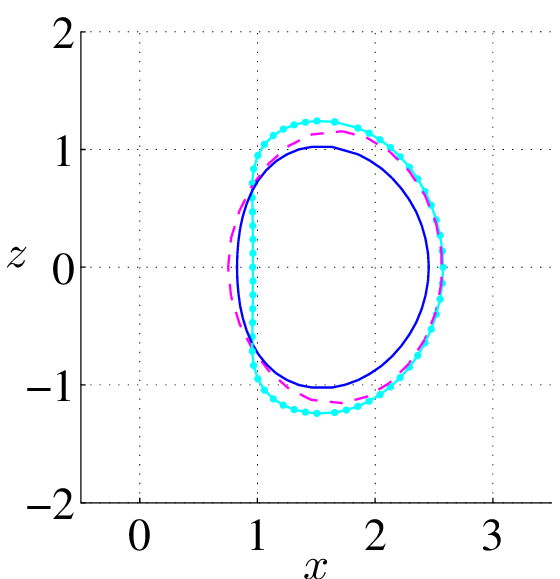}}
\put(-90,95){{\large (g)}}
\resizebox{4cm}{!}  {\includegraphics[width=\linewidth]{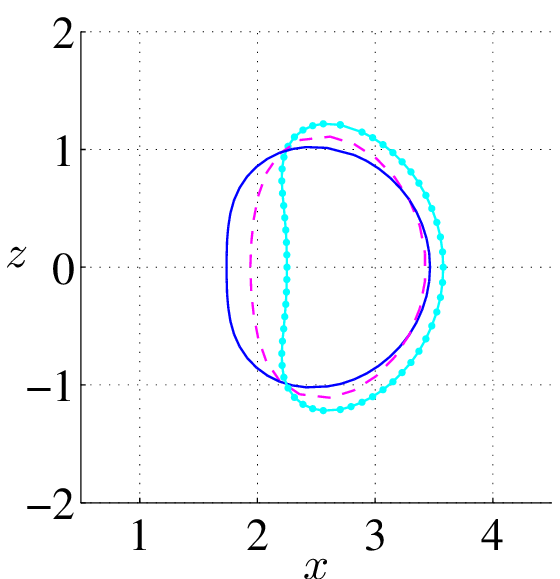}}
\put(-90,95){{\large (h)}}\\
\resizebox{4cm}{!}  {\includegraphics[width=\linewidth]{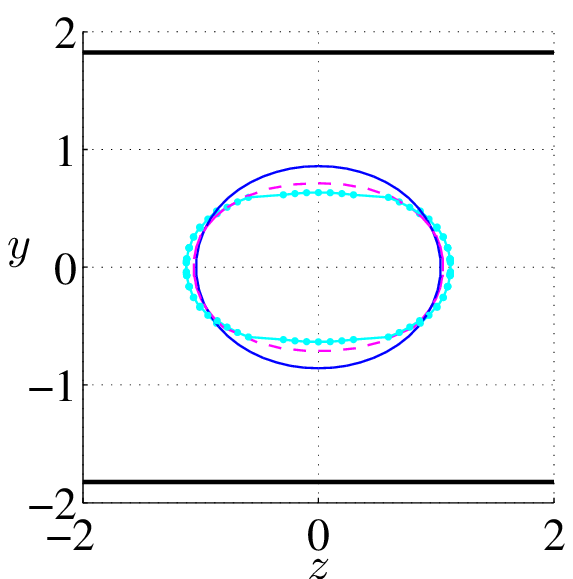}}
\put(-90,95){{\large (i)}}
\resizebox{4cm}{!}  {\includegraphics[width=\linewidth]{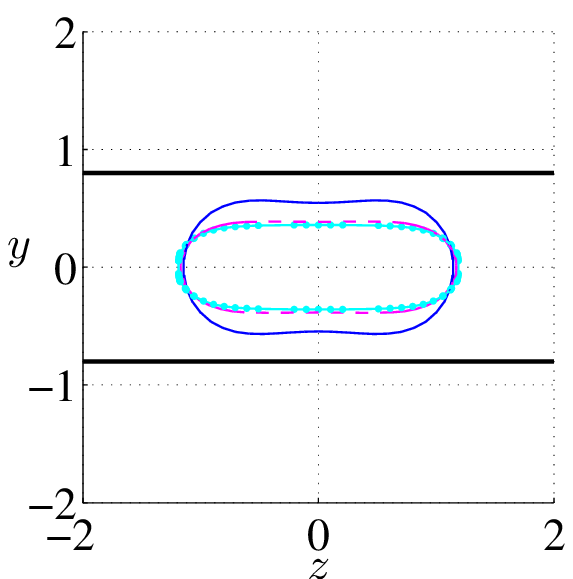}} 
\put(-90,95){{\large (j)}}
\resizebox{4cm}{!}  {\includegraphics[width=\linewidth]{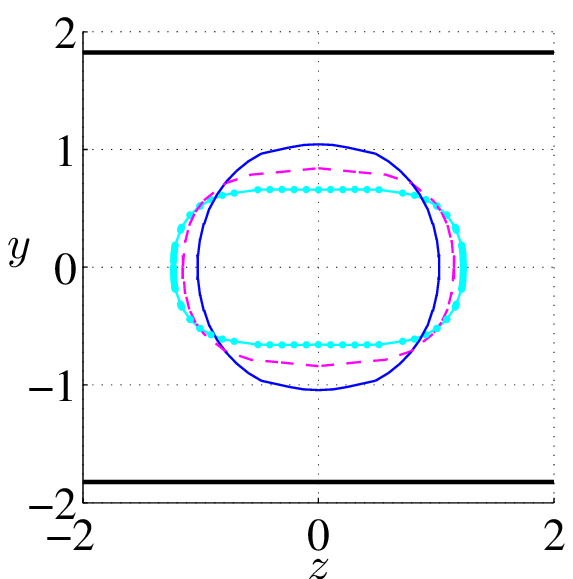}}
\put(-90,95){{\large (k)}}
\resizebox{4cm}{!}  {\includegraphics[width=\linewidth]{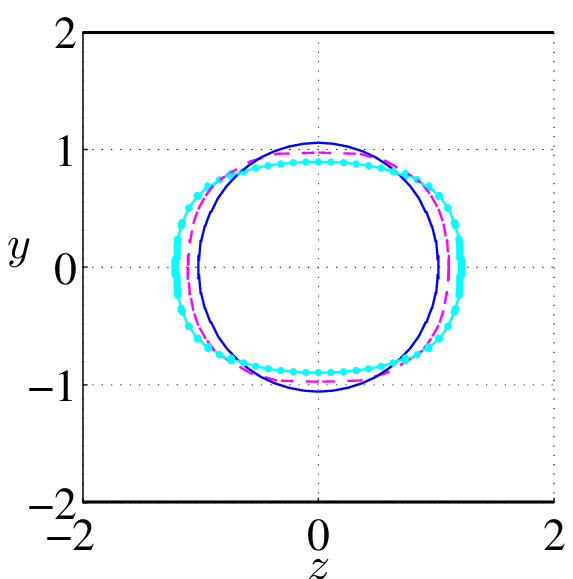}}
\put(-90,95){{\large (l)}}
\caption{Shape of the membrane on the $z=0$ (top row),  $y=0$ (mid row), and $x=0$ (bottom row) plane for a capsule (a),(e),(i) approaching the constriction, $x = -1.5$; (b),(f),(j) in the center of the constriction, $x = 0$; (c),(g),(k) exiting the constriction, $x = 1.5$;  and (d),(h),(l) after the constriction, $x = 2.5$, for the capillary numbers reported in the legend.}
\label{fig:behaviourcapsule}
\end{figure*}

\subsection{Motion in the constriction: Effect of the capillary number} \label{subsec:symmetric}
We first consider the sinusoidal symmetric constriction of length $l=4$, that is, geometry S4 in Table \ref{tab:channel_dimension_sym}. In Figure \ref{fig:behaviourcapsule} we show the capsule shape on the $z=0$ (first row), $y=0$ (second row), $x=0$ (third row) plane at different positions along its trajectory: before entering the constriction (first column), at the center of the constriction (second column), exiting the constriction (third column), and well past the constriction (fourth column). These profiles are cross-sections of the full 3D shape. We compare the behavior of 3 different capillary numbers.

When the capsule approaches the constriction its front part is affected by the accelerating and converging flow and adjusts to it through the elongated ``triangular'' shape seen in Figure \ref{fig:behaviourcapsule}(a). The capsule is stretched along the $x$-axis in the middle of the constriction  [Figure \ref{fig:behaviourcapsule}(b)], while is compressed along $x$, but stretched along $y$, when it exits from it [Figure \ref{fig:behaviourcapsule}(c)-(d)]. This ``rebound'' effect is caused by the rear of the capsule moving faster than the front in the diverging flow, a mechanism specular to that in a converging flow. It has been verified that the ``rebound'' effect occurs earlier for stiff capsules. Observe the concomitant stretching of the capsule in the $z$-direction [Figure \ref{fig:behaviourcapsule}(g)]. The steady state shape is  gradually recovered downstream [Figure \ref{fig:behaviourcapsule}(d),(h),(l)]. 

As expected, a soft capsule deforms more than a stiff one, and for capillary numbers high enough, it retains its characteristic croissant-like tail all along the trajectory. It is interesting to observe that the stiff capsule displays a nearly symmetric deformation in the middle of the constriction, see Figure \ref{fig:behaviourcapsule}(b). 

From these first considerations it emerges that the nonlinearity of the problem combined with the different response times of the membrane (associated to different capillary numbers), is responsible for a rich variety of dynamical behaviors that distinguish soft and stiff capsules. We try to characterize them quantitatively in what follows.
\begin{figure*}[!htbp]
\hspace{1.5cm}
\resizebox{7cm}{!}{\includegraphics[width=\linewidth]{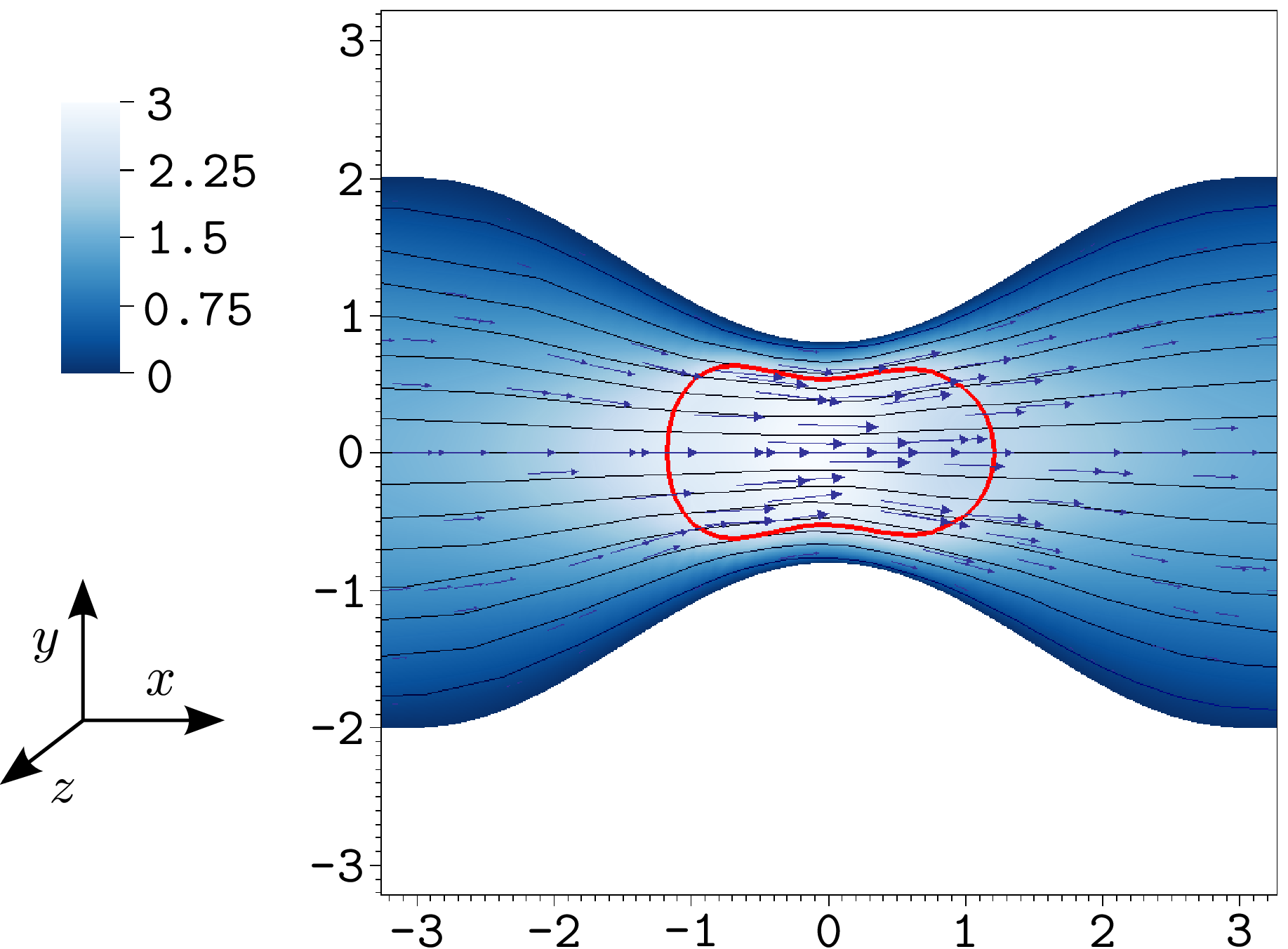}}
\put(-220,120){{\large (a)}}
\hspace{1.5cm}
\resizebox{7cm}{!}{\includegraphics[width=\linewidth]{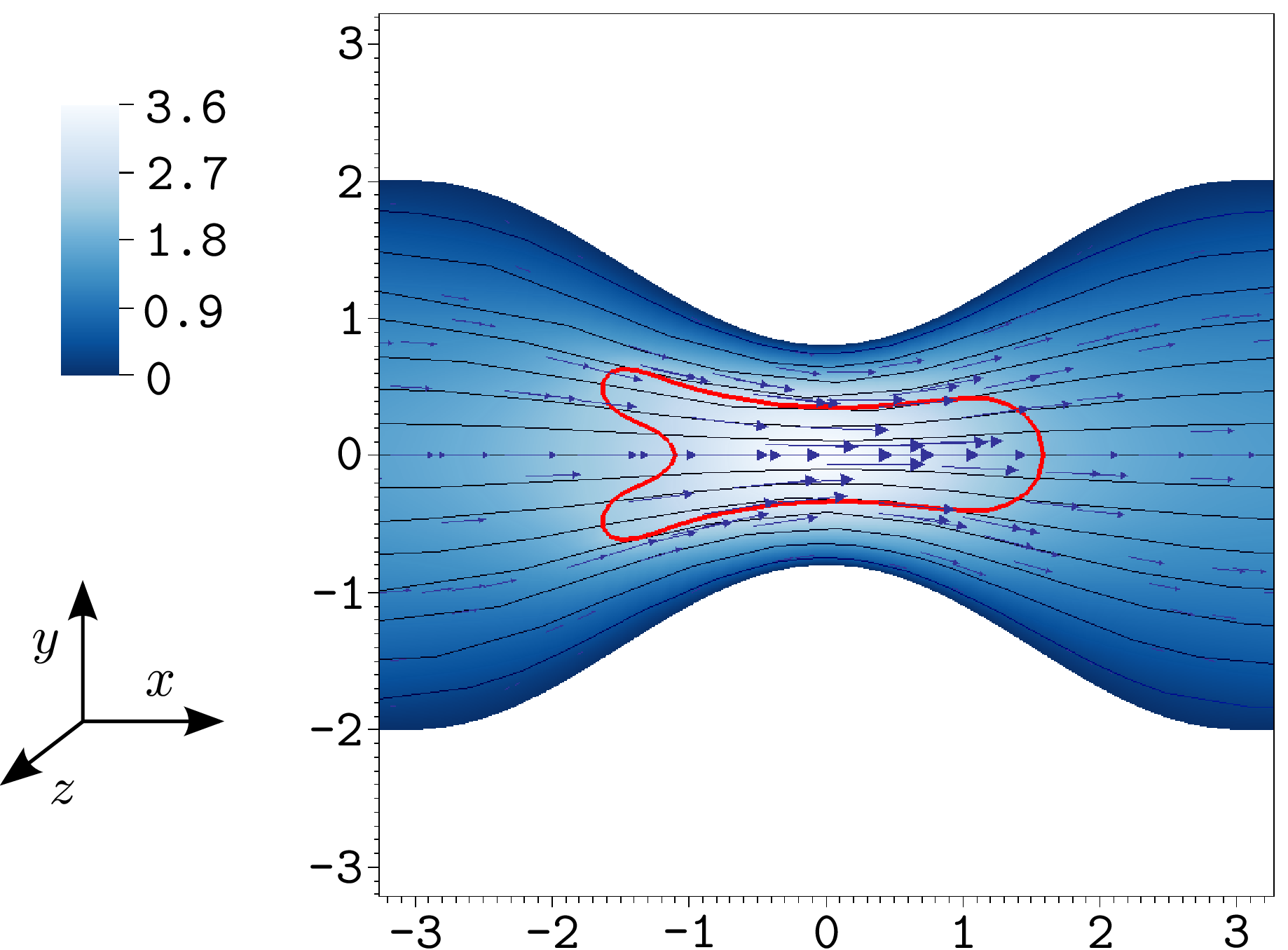}}
\put(-220,120){{\large (b)}}
\caption{Shape of the membrane (red) in the center of the constriction, for (a) $Ca=0.05$, and (b) $Ca=0.6$, the magnitude of the velocity (blue scale), the stream lines (black), and the velocity vector scaled by the velocity magnitude (dark blue) are displayed.}
\label{fig:flow}
\end{figure*}
\begin{figure*}[!htbp]
\begin{center}
\resizebox{7cm}{!}{\includegraphics[width=\linewidth]{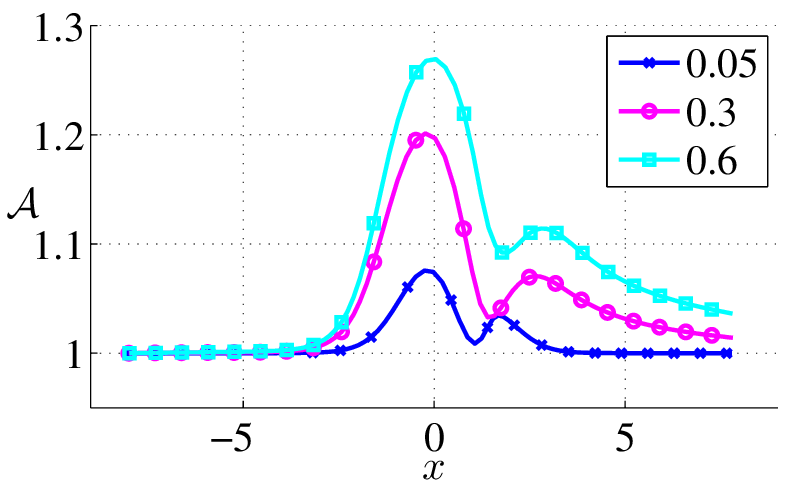}}
\put(-220,100){{\large (a)}}
\hspace{1.5cm}
\resizebox{7cm}{!}{\includegraphics[width=\linewidth]{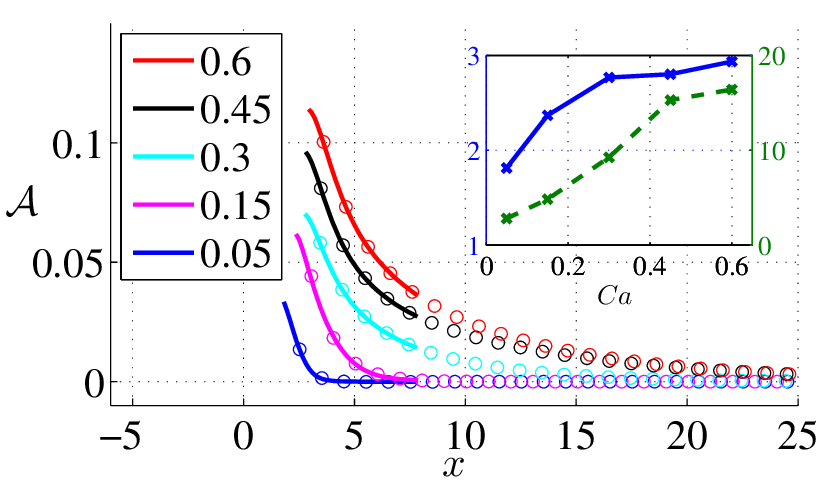}}
\put(-220,100){{\large (b)}}
\end{center}
\caption{(a) Evolution of the fractional change in area of the capsule, $\mathcal{A}$, for different capillary numbers as reported in the legend. (b) Exponential fitting (circles) to the fractional change in area decay toward the recovery of the steady state. We fit the curves to estimate where a 99\% recovery is achieved. The inset shows the  recovery position in $x$ versus Ca (in dash dark green) and the position of the second peak versus Ca (in solid blue).}
\label{fig:area_sym}
\end{figure*}
\begin{figure}[!htbp]
\begin{center}
\resizebox{6.6cm}{!}{\includegraphics[width=\linewidth]{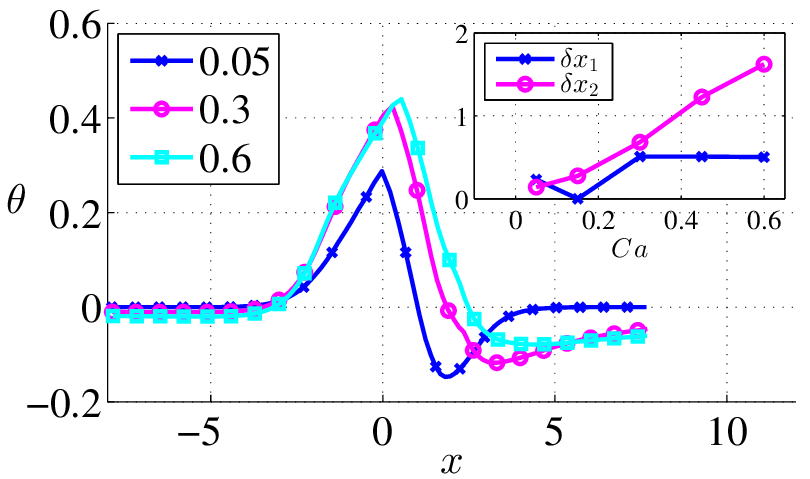}}
\put(-211,100){{\large (a)}}\\
\resizebox{6.8cm}{!}{\includegraphics[width=\linewidth]{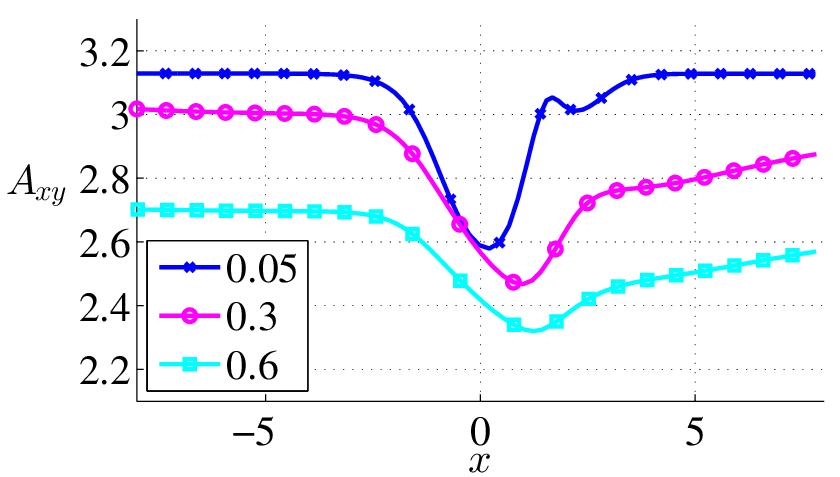}}
\put(-211,100){{\large (b)}}\\
\resizebox{6.6cm}{!}{\includegraphics[width=\linewidth]{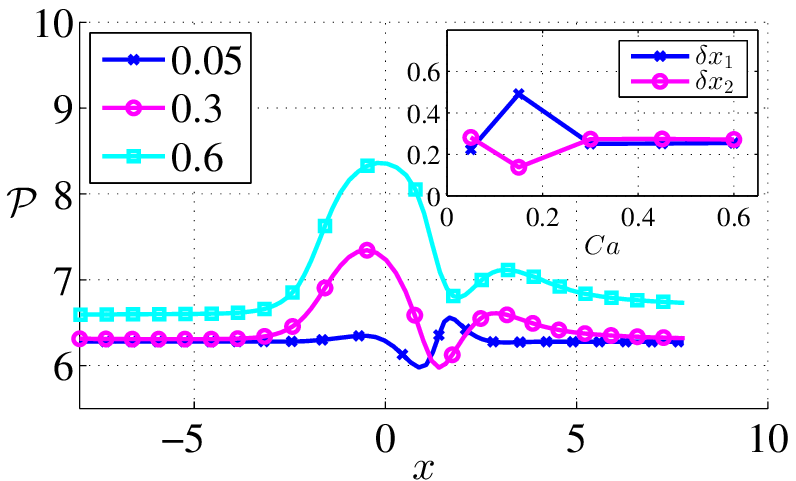}}
\put(-211,100){{\large (c)}}
\end{center}
\caption{\label{fig:deformation} Taylor parameter (a), area (b), and perimeter (c) of the capsule, measured on the cross section $z=0$ along the capsule trajectory for different capillary numbers. The inset of (a) shows the distance in $x$ between the first maxima of $\mathcal{A}$ and $\theta$, $\delta x_1$, and the second maximum of $\mathcal{A}$ and the minimum of $\theta$, $\delta x_2$, as a function of the capillary number. The inset of (c) shows the distance in $x$ between the first maxima, $\delta x_1$ and the second maxima, $\delta x_2$, of $\mathcal{A}$ and $\mathcal{P}$ as a function of the capillary number.}
\end{figure}

The velocity field around the stiffest and softest capsule when the capsule centroid is located about $x=0$ is reported in Figure \ref{fig:flow}. 
The maximum velocity is lower in the case of the stiff capsule, and the velocity profile is flattened inside it, tending to a ``solid body'' behavior. On the other hand the velocity distribution appears closer to the undisturbed flow in the case of the soft capsule, which expresses the fact that the capsule is more ``passively'' advected by the flow. The upstream-downstream symmetry of the flow appears to be broken more clearly for the stiff capsule.

The fractional change in area, $\mathcal{A}$, defined by
\begin{align}
\mathcal{A}(t) = \frac{A(t)}{A_0}, \label{equ:surfaceareaexpansion}
\end{align}
where $A(t)$ is the time-varying 
surface area of the capsule, is constant far away from the constriction as seen in Figure \ref{fig:area_sym}. 
This quantity starts increasing shortly before the capsule enters the constriction, reaches a maximum at $x=0$, and decreases before displaying a second local maximum which breaks the upstream-downstream symmetry.
The first maximum is reached about $x=0$ for all the capillary numbers since it is associated with the minimum width of the channel. However, the position of the second maximum is shifted to the right as the capillary number increases.
One may hypothesize that the second maximum is due to the above  mentioned rebound effect or stretching in the $y$-direction. However, we have verified that this hypothesis holds only for small capillary numbers.

We next examine the membrane relaxation past the constriction. The stiff capsule recovers the steady state configuration within our computational domain, while soft capsules ($Ca=0.3, 0.6$) do not. Nevertheless, we have estimated the $x$-position where a 99\% recovery of the steady state area is expected by fitting the decay of $\mathcal{A}$ from $x= x_{\mathrm{peak}_{2}, j}+0.7$ with a function $f:x \mapsto a_{j}\exp(b_{j}x)+c_{j}\exp(d_{j}x)$, where $j$ refers to the capillary number and $x_{\mathrm{peak}_{2}}$ indicates the position of the second peak, [see Figure \ref{fig:area_sym}(b)]. As shown in the inset of the figure, the relaxation time grows with $Ca$ and saturates for $Ca>0.45$.

Since it is difficult to measure the surface area variation experimentally, we estimate the deformation through different quantities more readily accessible from experimental data. These are the Taylor parameter, the perimeter and the area of the capsule on the $x-y$ plane. Our aim is to answer the question reported above: ``what is the most relevant quantity to measure to efficiently differentiate and sort cells by deformability?'', and to favor direct comparisons with experimental results.

Our visualizations reveal that despite the deformation along the $z$ axis is not negligible, it is about four times smaller than the deformation in the $x$ direction. Hence, it is acceptable to characterize the dynamics by analyzing the behavior on the $x-y$ plane. We attempt to quantify the anisotropy of the capsule shape on the $z=0$ plane by using the Taylor parameter: 
\begin{equation}
\theta=(L_x-L_y)/(L_y+L_x),
\end{equation}
where $L_x$ and $L_y$ are the sizes, in the $x$ and $y$ direction, of the rectangle in which the capsule can be inscribed. The parameter is zero if the capsule is perfectly circular. This quantity, plotted in Figure \ref{fig:deformation}(a), reveals that stiff capsules have a low degree of anisotropy in the constriction but at the same time undergo the largest rebound downstream (lowest negative value). This may sound counterintuitive since we expect stiff capsules to deform less, however the effect is clearly visible in  Figure \ref{fig:behaviourcapsule}(c) and (d).

The projected area of the capsule on the $z=0$ plane, A$_{xy}$, behaves markedly differently from $\mathcal{A}$, Figure \ref{fig:deformation}(b). It displays a minimum, whose exact position depends on $Ca$, about the center of the constriction. On the contrary, measurements of the perimeter, $\mathcal{P}$, on the same plane mimic quite reliably the behavior of $\mathcal{A}$, 
see Figure \ref{fig:deformation}(c). This is confirmed by the inset in Figure \ref{fig:deformation}(c), where the positions of the two maxima for $\mathcal{P}$ and $\mathcal{A}$ (the one in $x=0$ and the second peak) are compared by defining the parameter $\delta x_j=|x(\max_j \mathcal{A})-x(\max_j \mathcal{P})|$, where $j=1,2$ refers to the first and second maximum. We also note that there is not such a consistent correspondence between the maxima of $\mathcal{A}$ and those of the Taylor parameter: in the inset of Figure \ref{fig:deformation}(a), we show $\delta x_1=|x(\max_1 \mathcal{A})-x(\max \theta)|$, and $\delta x_2=|x(\max_2 \mathcal{A})-x(\min \theta)|$, since the variable $\theta$ does not have a second maximum but displays a minimum instead.

\begin{figure}[!htbp]
\centering
  \resizebox{6.5cm}{!}{\includegraphics[width=\linewidth]{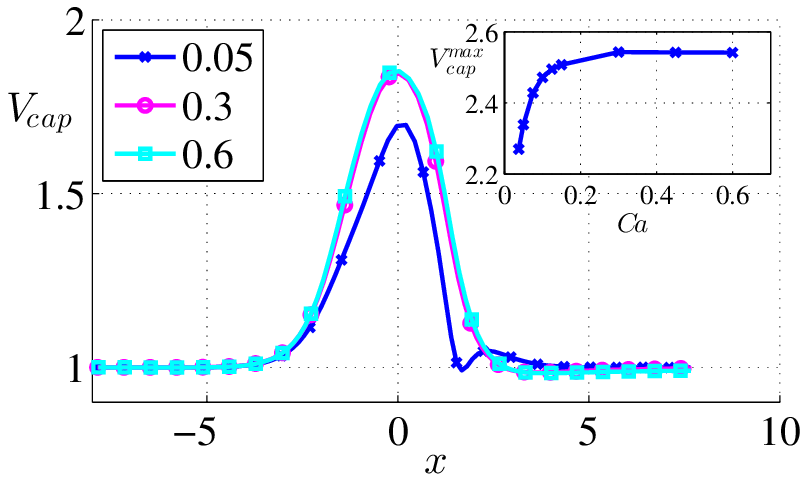}}
  \put(-217,100){{\large (a)}}\\
  \resizebox{6.5cm}{!}{\includegraphics[width=\linewidth]{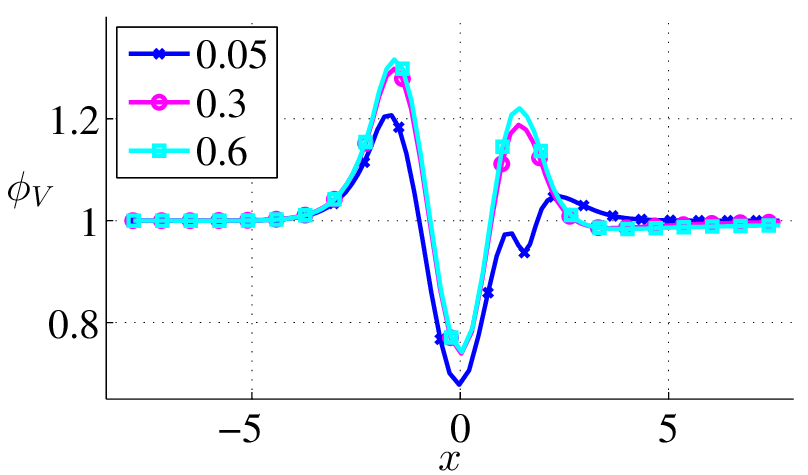}}
  \put(-217,100){{\large (b)}}
\caption{(a) Velocity of the capsule rescaled by its initial value, $V_{cap}$,  as a function of the capsule centroid position for different capillary numbers; the inset shows the maximum velocity versus the capillary number; (b) velocity of the capsule rescaled by the local flow velocity, $\phi_{V}$, for different capillary numbers as reported in the legend.}
\label{fig:velocity_sym}
\end{figure}
\begin{figure}[!htbp]
\centering
\resizebox{7.5cm}{!}{\includegraphics[width=\linewidth]{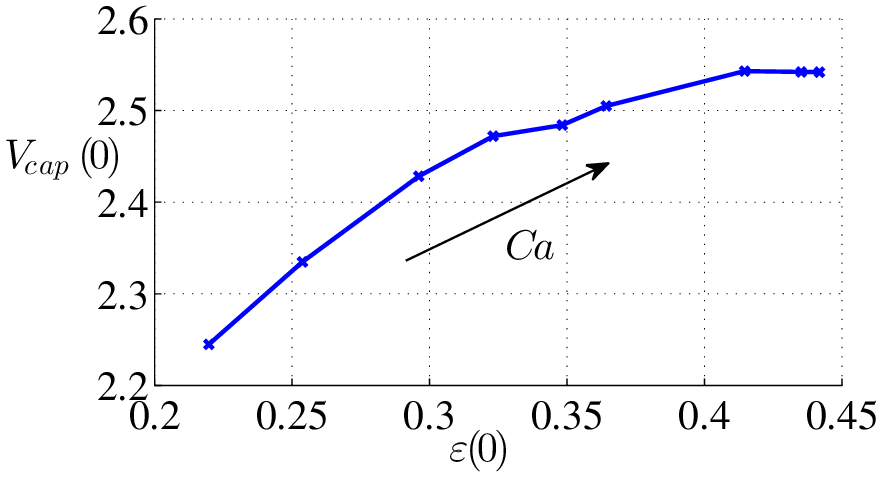}}
\caption{Velocity of the capsule, $V_{cap}(0)$, as a function of the minimum distance from the wall, $\varepsilon(0)$ measured at the center of the constriction.
\label{fig:behaviour3}}
\end{figure}
\begin{figure}[!htbp]
\centering
\resizebox{6.5cm}{!}{\includegraphics[width=\linewidth]{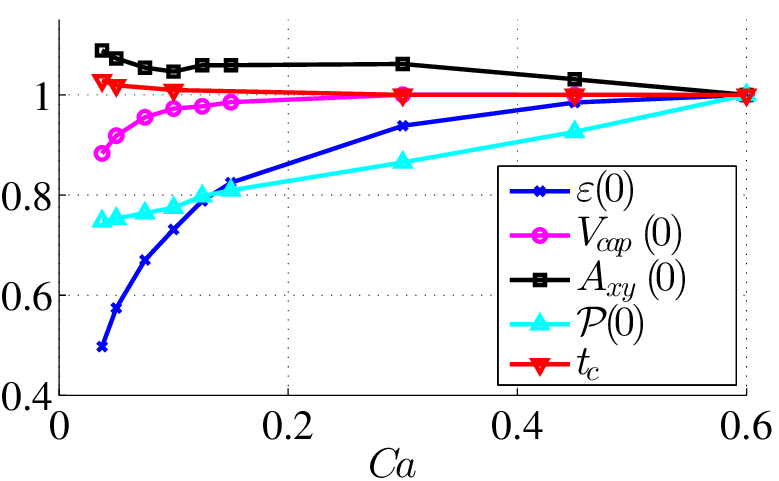}}
\caption{Summary of the results obtained so far for the quantities listed in the legend, measured at the center of the constriction, plotted as a function of the capillary number, and rescaled by their value at $Ca=0.6$.}
\label{fig:behaviour}
\end{figure}

Next, the velocity of the capsule, $V_{cap}$, is examined.
Since in these simulations the capsule is located on the channel midline, there is no migration velocity in the $y$ and $z$ direction. The capsule moves faster than the cross-section averaged flow velocity ($V_{cap}$>1), accelerating and decelerating
with the flow inside the constriction, see Figure \ref{fig:velocity_sym}(a).
 The velocity is maximum in $x=0$. The softer is the capsule, the faster it is in the constriction. This might be simply explained by the fact that soft capsules stretch more around the mid-plane of the channel (Figure \ref{fig:behaviourcapsule}), thereby, given the background fluid velocity profile, their membrane is advected by larger velocities. 
 The velocity decreases when the capsule leaves the constriction. 

Observe that despite capsules with $Ca=0.3$ and $Ca=0.6$ deform quite differently (Figure \ref{fig:behaviourcapsule}, \ref{fig:area_sym}), their velocity is very similar  [Figure \ref{fig:velocity_sym}(a)]. This can be possibly explained by the fact that the velocity is related to
the minimum distance from the wall, $\varepsilon$, which occurs at the  front for both capsules. The front shape is very similar for high capillary numbers, in fact, the difference in deformation between $Ca=0.3$ and $Ca=0.6$ is localized on the rear part which develops the typical croissant-like tail. The behavior of the maximum velocity as a function of the capillary number is shown in the inset of Figure \ref{fig:velocity_sym}(a). The steep increase for small capillary numbers is followed by a plateau for $Ca>0.3$.

We observe that for the stiffest capsule ($Ca=0.05$), a second maximum appears shortly after the first one, suggesting that a qualitatively different behaviour distinguishes stiff from soft capsules. This difference is more clearly observed if the velocity of the capsule is rescaled by the local average flow velocity $V_f(x)$:
\begin{align}
\phi_{V} = \frac{V_{cap}}{V_{f}}. \label{equ:scaled_velocity}
\end{align}
The ratio $\phi_V$ is displayed in
figure \ref{fig:velocity_sym}(b): it is maximum immediately before and after the constriction and minimum at its center. In fact, although the capsule is stretched in this location, it occupies relatively more space in the constriction than in the straight channel, the distance between the wall and the capsule membrane is smaller, and the lubrication force is larger. As a result of its limited deformability and quick response time, the stiffest capsule ($Ca=0.05$) relaxes and stretches in $y$ when it is still exiting the constriction; this causes its membrane to approach the diverging walls and slow down with respect to the flow, since $\varepsilon$ suddenly decreases. This possibly explains the local minimum observed in Figure \ref{fig:velocity_sym}(b).

In Figure \ref{fig:behaviour3} we check how the capsule velocity is related to the minimum distance, $\varepsilon$, between the membrane and the wall of the constriction. The data are taken at the center of the constriction. The dependence is not linear, as hypothesized in \cite{Park13}.  Less deformed capsules get closer to the wall than soft capsules.
\begin{figure}[!htbp]
\centering
\resizebox{6.5cm}{!}{\includegraphics[width=\linewidth]{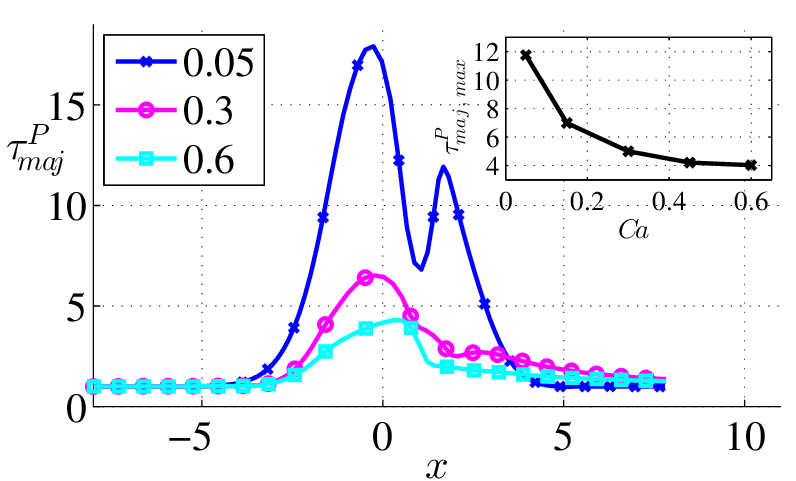}}
\put(-200,100){{\large (a)}}\\
\resizebox{6.5cm}{!}{\includegraphics[width=\linewidth]{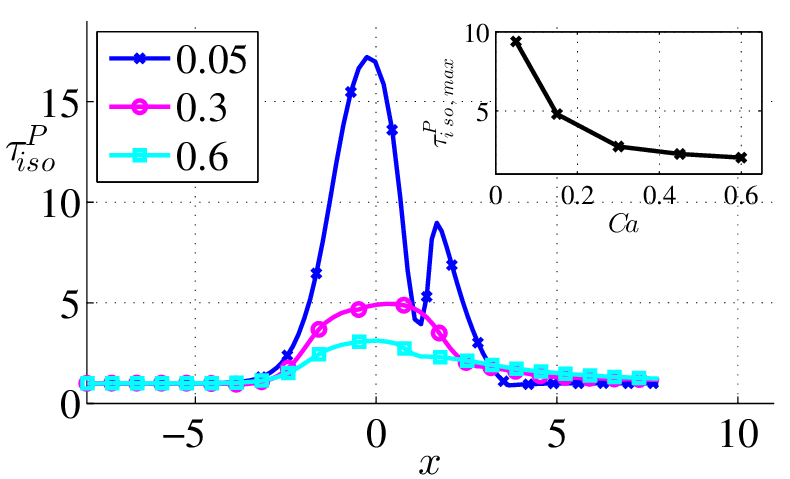}}
\put(-200,100){{\large (b)}}
\caption{Evolution of (a) the principal major stress, $\tau^P_{maj}$, and (b) the principal isotropic stress, $\tau^P_{iso}$, rescaled by their initial value, for the capillary numbers reported in the legend. The inset shows their maximum value versus the capillary number.}
\label{fig:stress_sym}
\end{figure}

We summarize the results obtained so far in Figure \ref{fig:behaviour}, where the values at $x=0$ of the minimum distance from the wall, $\varepsilon$, the velocity of the centroid of the capsule, and the area and the perimeter on the $z=0$ plane are plotted as a function of the capillary number and are compared with the retention time $t_c$ inside the device measured from $x=-8$ (at the release time) to $x=8$. All the quantities are rescaled by their values at $Ca=0.6$. In relative terms, $\varepsilon$ displays the greatest variation between stiff and soft capsules, however, it plateaus for large $Ca$ making it difficult to distinguish the behavior in this range. On the other hand, the perimeter  grows linearly for the entire capillary number range, which makes it the most convenient parameter to distinguish capsules by deformability. The retention time of the capsule, $t_c$, decreases as the capillary number increases by about a 3\%, which is expected given the larger velocities of soft capsules. This fact may be exploited to distinguish capsules by deformability at low capillary numbers if the difference is amplified, for example, by an array of constrictions.

We observe that, in general, area-deformation correlated quantities ($\mathcal{P}$)
grow linearly with $Ca$, while velocity-correlated quantities ($V_{cap}$, $\varepsilon$, $t_c$) saturate for large $Ca$ but display a steeper increase for small $Ca$. When comparing the quantities in absolute terms we see that $\varepsilon$ varies between $0.22$ to $0.44$ ($100\%$ increase) while $\mathcal{P}$ between $6.25$ to $8.35$ ($33\%$ increase).

\begin{figure}[!htbp]
\centering
\resizebox{6.5cm}{!}{\includegraphics[width=\linewidth]{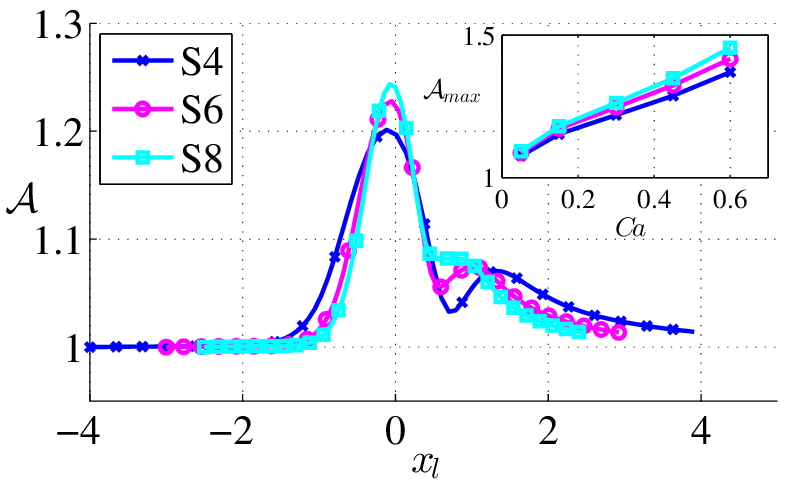}}
\put(-200,100){{\large (a)}}\\
\resizebox{6.5cm}{!}{\includegraphics[width=\linewidth]{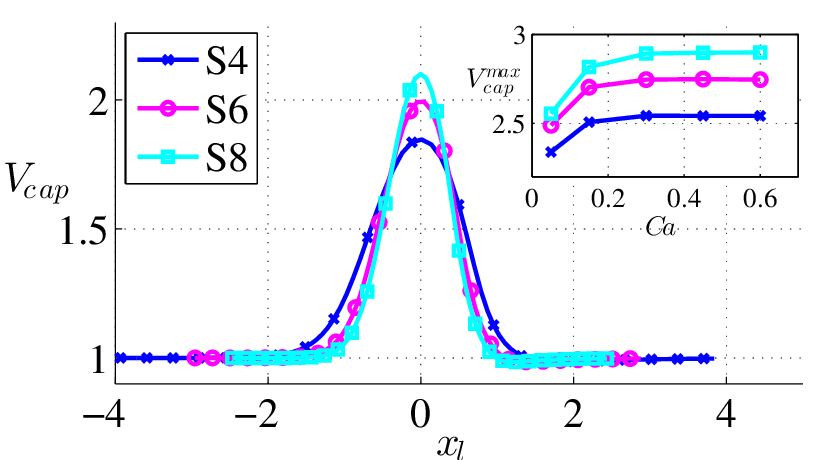}}
\put(-200,100){{\large (b)}}
\caption{(a) Evolution of the fractional change in area for a capsule with $Ca=0.3$, and, as an inset, maximum of the fractional change in area versus the capillary number for different constriction lengths as reported in the legends. (b) Evolution for $Ca=0.3$, of the capsule centroid velocity, $V_{cap}$, and as an inset, maximum of the capsule velocity versus the capillary number for the constriction lengths reported in the legend.}
\label{fig:geo_sym}
\end{figure}

We finally analyse the stress on the capsule membrane along the constricted channel.
In Figure \ref{fig:stress_sym} we plot the major principal tension: $$\tau^P_{maj}(t)=\max_{\boldsymbol{x}}\left[\tau^P_1(\boldsymbol{x},t),\tau^P_2(\boldsymbol{x},t)\right]$$ and the isotropic principal tension\\
 $$\tau^{P}_{iso}(t) =\max_{\boldsymbol{x}}\left[\tau^P_1(\boldsymbol{x},t)+\tau^P_2(\boldsymbol{x},t)\right]/2$$
divided by their steady state values. 

Opposite to the behavior of the fractional change in area, $\mathcal{A}$, the stress of soft capsules is lower, in relative terms, as shown both by $\tau^P_{maj}$ and $\tau^P_{iso}$. The stiffest capsule ($Ca=0.05$), which has a steady state shape very close to a sphere, has an initial principal isometric and principal major stress very close to zero. Moreover, we can clearly notice that soft capsules take a longer time to recover their initial stress. The inset shows that the dimensionless major and isotropic tensions ($\tau^P_{maj}/G_s$, $\tau^P_{iso}/G_s$) at  $x=0$ decrease with the capillary number.
\begin{figure}[!htbp]
\centering
\resizebox{6.5cm}{!}{\includegraphics[width=\linewidth]{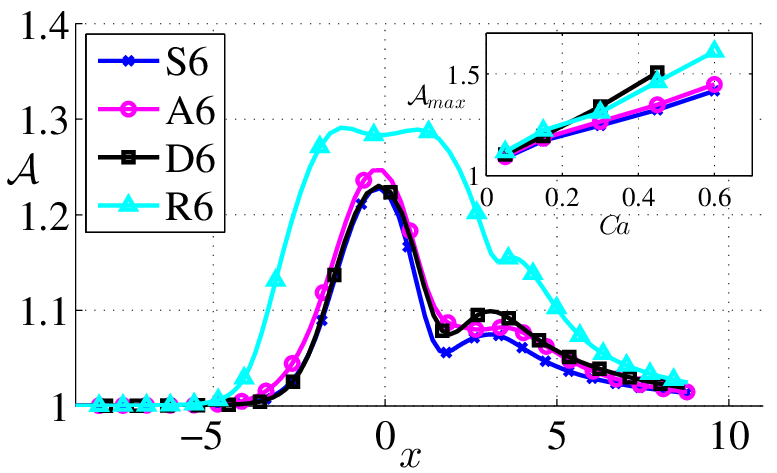}}
\put(-200,100){{\large (a)}}\\
\resizebox{6.5cm}{!}{\includegraphics[width=\linewidth]{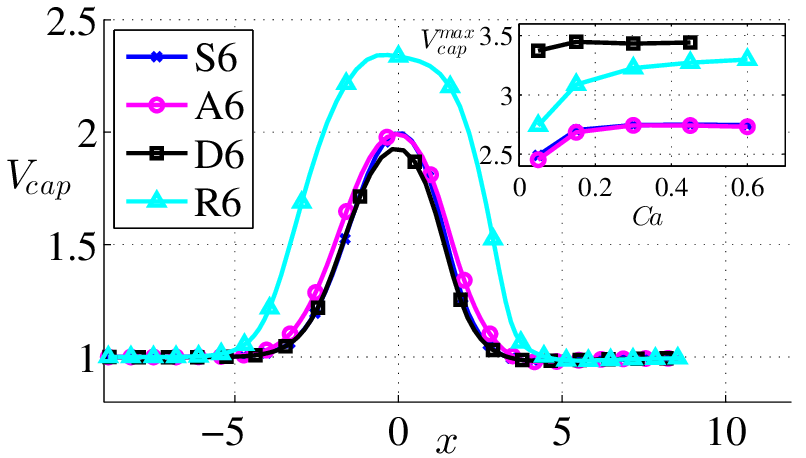}}
\put(-200,100){{\large (b)}}
\caption{(a) Evolution, for $Ca=0.3$, of the fractional change in the area, $\mathcal{A}$ and as inset, the maximum of the fractional change in the area versus the capillary number for different geometries as reported in the legend. (b) Evolution for $Ca=0.3$, of the capsule centroid velocity, $V_{cap}$, rescaled by its initial value and as inset, the maximum velocity versus Ca, for different geometries as reported in the legend.}
\label{fig:geo}
\end{figure}
\subsection{Effect of the constriction geometry} \label{subsec:length} 
Finally, the sensitivity of the capsule transient dynamics to the constriction geometry is investigated by studying the capsule deformation and the variation of the centroid velocity in different geometries. 

First, the effect of varying the constriction length, $l$, is considered by studying the motion in two longer configurations, S6 and S8 (see Table \ref{tab:channel_dimension_sym}). 
It emerges that the evolution of the fractional change in area, $\mathcal{A}$, is affected by $l$, [Figure \ref{fig:geo_sym}(a)]. In order to directly compare the different cases, the deformation is plotted as a function of a dimensionless coordinate, $x_l=2x/l$ so that the start and end of the constriction are located at $x_l=\pm1$. The deformations are larger in S6 and S8 than in S4 although the flow acceleration/deceleration is smoother in the longer geometries. For $l$ large enough,
 we expect the capsule to reach a new steady state inside the constriction, characterized by larger deformations.  
In longer constrictions, the minimum and the second maximum of $\mathcal{A}$ are less pronounced, 
likely because the section area varies more gradually. 
Note also that in the geometry S4 the capsule seems to feel the effect of the constriction from an earlier to a later stage, probably because the acceleration in the converging part is stronger and the time spent in the constriction is shorter than the membrane relaxation times.

The inset in Figure \ref{fig:geo_sym}(a), which displays the maximum of the fractional change in area versus the capillary number, confirms greater deformations for longer constrictions. It also shows that the differences in deformation are more evident for soft capsules. It is interesting to note the similarities between the inset of Figure \ref{fig:geo_sym}(a) and Figure 6 in \cite{Kaoui09}, where the capillary number is plotted versus the vesicle excess area. Since stiff capsules deform more quickly, one would be tempted to say that the stiffest capsule ($Ca=0.05$) reaches an equilibrium shape even in the shortest constriction S4. In fact, its deformation seems to be independent of the constriction length. 
Interestingly, this is not the same for the velocity, which differs at small $Ca$ [Figure \ref{fig:geo_sym}(b) inset]. Hence, apparently negligible deformations may be rather and more simply attributed to the stiffness. 

The behavior of the capsule velocity, $V_{cap}$, 
depends on the length of the constriction as seen in Figure \ref{fig:geo_sym}(b). The capsule velocity is larger in longer constrictions because the membrane has more time to deform and localize in the centre of the channel. The inset in Figure \ref{fig:geo_sym}(b), which displays the maximum of the velocity, $V_{cap}^{max}$, versus the capillary number for different constriction lengths, shows that the maximum velocity  increases with the capillary number from $0.05$ to $0.3$ and is roughly constant for capillary numbers between $0.3$ and $0.6$; the maximum velocity is higher if the constriction is longer, and the same qualitative behavior is observed as a function of the capillary number for the three different geometries.

We finally analyse the effect of different constriction shapes: an asymmetrical constriction, A6, a rectangular constriction, R6, characterized by a sharper reduction of the cross section, and the flow in a constricted square duct, D6, as reported in Table \ref{tab:channel_dimension_sym} and Fig. \ref{fig:sketch}. 
Figure \ref{fig:geo} displays the evolution of the fractional change in area and the velocity for the
 different geometries compared to the reference case S6. 
In the asymmetric constriction A6, the capsule is displaced from the mid plane of the channel, since it has to pass through the 'chicane' shaped by the asymmetry, and it stretches slightly more than in a symmetrical constriction of the same length. We can thus conclude that asymmetrical constrictions are more efficient in deforming capsules than symmetrical ones. The difference in deformation between A6 and S6 over the range of capillary numbers is confirmed by the inset of Figure \ref{fig:geo}(a). Interestingly, however, the velocity of the capsule in the asymmetrical constriction does not differ much from that of the reference case S6 as shown by the inset in Figure \ref{fig:geo}(b). 

The confinement in a square duct does not appear to considerably affect the dynamics in relative terms, in fact, both the fractional change in the area and the velocity of the capsule (scaled by their initial value) are similar for case S6 and D6, cf.\ Figure \ref{fig:geo}(a)-(b).
We notice however small differences in the second maximum of the area and the velocity. As a result of the confinement, the capsule is not free to deform in the z direction (geometry D6), and the flow is more affected by the presence of the capsule than for the case in a quasi-twodimensional geometry. Indeed, the non-rescaled quantities reported in the insets of Figure \ref{fig:geo}(a)-(b) differ considerably between S6 and D6. The velocity in the square duct is higher, due to the confinement.

Remarkable differences are observed also for the rescaled quantities when comparing the behavior in the rectangular constriction R6: the deformation is larger and experienced for longer times; the capsule moves faster as noted earlier for the case of longer constrictions. These effects can be explained by the fact that the capsule is exposed to the faster flow in the constriction over a longer distance.


%

\section{Conclusion and outlook}
We have studied the motion of a deformable capsule through a constricted microchannel and quantified, in particular, how the capsule deformation, velocity, retention time, and the maximum stress of the membrane are affected by the capillary number and the constriction shape. The capsule stress-free configuration is a sphere of unit radius. Simulations are performed for capillary numbers between $Ca = 0.05$ and $Ca = 0.6$.
Our study is motivated in the context of microfluidic devices and is relevant to applications in medical diagnostic that involve mechanical characterization and sorting of cell samples.

We first calculate the equilibrium shape of a capsule in an infinitely long channel with spanwise periodic or no-slip boundary conditions and use it as initial condition for calculations in constricted channels. This guarantees that the dynamics of the capsule is solely affected by the geometry of the domain, instead of being also influenced by the convergence to a steady state. The steady state is nearly spherical for stiff capsules, whereas it develops a front-rear asymmetry and displays first a bullet-like, and then a croissant-like shape by increasing the capillary number.
The stress is larger on the front part, and this is where rupture may occur. These results confirm the findings of previous studies. 

 The capsule moves faster than the cross-section average flow velocity accelerating and decelerating with the flow inside the constriction. The velocity is maximum at the center of the constriction. The softer is the capsule, the faster it is in the constriction.

The nonlinearity of the problem combined with the different response times of the membrane (defined by the capillary numbers), is responsible for a rich variety of dynamical behaviors. To mention some of them:  ({\it i}) the variation of the surface area and maximum stresses as the capsule moves through the constriction are not symmetric with respect to the constriction (streamwise) mid point $x=0$, rather a second peak appears past the constriction whose location depends on the $Ca$ number; the larger is $Ca$, farther downstream the peak is located; ({\it ii}) a similar dependence on the capillary number is observed in the relaxation time necessary for the capsule to restore its steady state past the constriction; ({\it iii}) the behavior of the velocity of the capsule centroid along the capsule trajectory displays a qualitatively different behavior for small $Ca$, which consists in the appearance of a second peak past the constriction.

We estimate the deformation by measuring the variation of the three-dimensional surface area and a series of alternative quantities that we imagine easier to extract from experiments. These are the Taylor parameter, the perimeter and the area of the capsule in the $x-y$ plane. Our aim is to identify an observable to measure experimentally to efficiently distinguish capsules by deformability. 
We report that the perimeter is the quantity that reproduces the behavior of the three-dimensional fractional change in area the best, followed by the Taylor parameter, while the area of the capsule in the $x-y$ plane has a markedly different behavior. Hence, in conclusion, if the variation of the surface area cannot be easily measured, the perimeter in the $x-y$ plane is a valid alternative.

We find that the velocity at the center of the constriction and the retention time are adequate parameters to distinguish capsules with ``low'' capillary numbers ($0.05 \lesssim Ca \lesssim 0.3$), while large capillary numbers are more clearly distinguished by the deformability and perimeter on the $x-y$ plane at the center of the constriction. This is because the velocity plateaus for $Ca \gtrsim 0.3$ while the maximum deformation still grows linearly. The reason of this different behavior is the fact that the velocity is related to the minimum distance from the wall, which stabilizes for high capillary numbers, while the deformation grows linearly in the entire range of our calculations, and for high capillary numbers it is associated to the transition from the convex to the concave shape of the capsule rear part and the appearance of the typical ``croissant-like'' shape.  On the other hand, the minimum distance from the wall is always measured at the front for soft capsules, and the front shape is very similar for high capillary numbers, all the changes occurring on the rear part.

Finally, we have observed that longer constrictions and $z$-confined (versus $z$-periodic) domains cause larger deformations and velocities. Interestingly, if the deformation and velocity in $z$-confined domains are rescaled by the equilibrium shape deformation and velocity, their behavior is undistinguishable from that in a $z$-periodic domain. In contrast, a remarkably different behavior is reported in sinusoidal and smoothed rectangular constrictions indicating that the capsule behavior is particularly sensitive to abrupt changes in the cross section. In a smoothed rectangular constriction larger deformations and velocities occur over a larger distance.

We hope that our numerical calculations will serve as a reference for future experiments, to validate the model of membrane dynamics and the effects of the viscosity in the membrane model. 
 
\section*{Acknowledgements}

Computer time provided by SNIC (Swedish National Infrastructure
for Computing) is gratefully acknowledged. This work was partially supported by the European Research
Council Grant No.\ ERC-2013-CoG-616186, TRITOS. CR acknowledges financial support from the G\"oran Gustaffson Foundation. LZ acknowledges the finalcial support from the European Research Council grant (ERC simcomics
– 280117) for his stay at EPFL. We all thank Prof. Dhrubaditya Mitra from Nordita and Prof. Debjani Paul from IIT Mumbai for the insightful discussions.

\bibliography{Article.bbl}

\end{document}